\documentclass[aps,prmaterials,twocolumn,longbibliography,nobibnotes,showkeys]{revtex4-2}

\usepackage{graphicx}% Include figure files
\usepackage{dcolumn}% Align table columns on decimal point
\usepackage{bm}% bold math
\usepackage[version=3]{mhchem}
\usepackage{color}

\begin{document}
\title{Tuning the valence and concentration of europium and luminescence centers in GaN through co-doping and defect association}
\author{Khang Hoang}
\email{khang.hoang@ndsu.edu}
\affiliation{Center for Computationally Assisted Science and Technology \& Department of Physics, North Dakota State University, Fargo, North Dakota 58108, United States}

\date{\today}

\begin{abstract}

Defect physics of europium (Eu) doped GaN is investigated using first-principles hybrid density-functional defect calculations. This includes the interaction between the rare-earth dopant and native point defects (Ga and N vacancies) and other impurities (O, Si, C, H, and Mg) unintentionally present or intentionally incorporated into the host material. While the trivalent Eu$^{3+}$ ion is often found to be predominant when Eu is incorporated at the Ga site in wurtzite GaN, the divalent Eu$^{2+}$ is also stable and found to be predominant in a small range of Fermi-level values in the band-gap region. The Eu$^{2+}$/Eu$^{3+}$ ratio can be tuned by tuning the position of Fermi level and through defect association. We find co-doping with oxygen can facilitate the incorporation of Eu into the lattice. The unassociated Eu$_{\rm Ga}$ is an electrically and optically active defect center and its behavior is profoundly impacted by local defect--defect interaction. Defect complexes such as Eu$_{\rm Ga}$-O$_{\rm N}$, Eu$_{\rm Ga}$-Si$_{\rm Ga}$, Eu$_{\rm Ga}$-H$_i$, Eu$_{\rm Ga}$-Mg$_{\rm Ga}$, and Eu$_{\rm Ga}$-O$_{\rm N}$-Mg$_{\rm Ga}$ can efficiently act as deep carrier traps and mediate energy transfer from the host into the Eu$^{3+}$ $4f$-electron core which then leads to sharp red intra-$f$ luminescence. Eu-related defects can also give rise to defect-to-band luminescence. The unassociated Eu$_{\rm Ga}$, for example, is identified as a possible source of the broad blue emission observed in n-type, Eu$^{2+}$-containing GaN. This work calls for a re-assessment of certain assumptions regarding specific defect configurations previously made for Eu-doped GaN and further investigation into the origin of the photoluminescence hysteresis observed in (Eu,Mg)-doped samples. 

\end{abstract}

\pacs{}
%\keywords{}

\maketitle


\section{Introduction}\label{sec;intro}

\begin{figure}[t]%
\vspace{0.2cm}
\includegraphics*[width=0.95\linewidth]{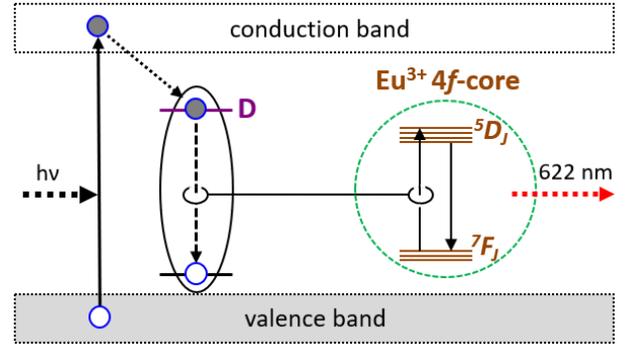}
\caption{Schematic illustration of non-resonant Eu$^{3+}$ excitation in GaN. Following a band-to-band excitation of the host, an electron is excited from the valence band to the conduction band. The excited electron is then trapped at the defect level D before recombining non-radiatively with a hole and the recombination energy is transferred into the Eu$^{3+}$ $4f$-electron core. A mechanism involving hole trapping is similar.}
\label{fig;excitation}
\end{figure}

Rare-earth (RE) doped III-nitrides are of interest for optoelectronic and spintronic applications \cite{ODonnell2010Book}. Thanks to their $4f$-electron core, which is well shielded by the outer $5s^2$ and $5p^6$ electron shells, these RE dopants offer very sharp intra-$f$ optical transitions at wavelengths from the infrared to ultraviolet. GaN doped with trivalent europium (Eu$^{3+}$), for example, emits visible light in the red spectral region and is considered as a promising candidate for light-emitting diodes (LEDs)~\cite{Fujiwara2014JJAP,Mitchell2018JAP}. In general, a RE luminescence center can be optically excited by resonant (direct) or non-resonant (indirect) excitation. In the former the excitation energy is directly absorbed into the $4f$-electron core, whereas in the latter it is indirectly transferred from the host. The non-resonant excitation mechanism is believed to be mediated by defects which act as carrier traps; see Fig.~\ref{fig;excitation}. An electron (hole) trapped at a defect level can then recombine non-radiatively with a hole (electron) from the valence (conduction) band or some acceptor (donor) level, and the recombination energy is transferred into the $4f$-core. RE-related defects are of interest in particular because of the close proximity of the RE ion to the carrier trap which enhances energy transfer efficiency. In addition to the intra-$f$ luminescence, RE-related defects, like other defects in a semiconductor host, can also act as carrier traps for defect-to-band optical transitions which do not involve energy transfer into the $4f$-core. A detailed understanding of defect physics in RE-doped semiconductors is thus essential to understanding their properties and to designing materials with improved performance. 

Experimentally, while the trivalent Eu$^{3+}$ ion was found to be predominant in Eu-doped GaN samples and multiple Eu$^{3+}$ luminescence centers were observed~\cite{Fleischman2009APB,Roqan2010PRB,Lorenz2010APL,ODonnell2011OM,Woodward2011OM,Woodward2011APL,Wakamatsu2013JAP,Mitchell2014JAP,Timmerman2020PRA}, the divalent Eu$^{2+}$ has also been found or suspected to be present \cite{Maruyama2002JCG,Hashimoto2003JJAP,Tanaka2003PSSC,Kim2004JAP,Hite2006APL,Mahalingam2007AFM,Mitchell2017MCP,Nunokawa2020JAP}. In addition to being of interest for its magnetic properties~\cite{Hashimoto2003JJAP,Tanaka2003PSSC,Nunokawa2020JAP}, Eu$^{2+}$ can offer useful luminescence centers in its own right. Mahalingam et al.~\cite{Mahalingam2007AFM}, for example, attributed the broad blue emission in Eu-doped GaN/SiO$_2$ nanocomposites to the presence of Eu$^{2+}$. Mitchell et al.~\cite{Mitchell2017MCP} reported a thorough work on the control of the Eu$^{2+}$/Eu$^{3+}$ ratio in GaN through co-doping and by tuning the growth conditions and were able to achieve high Eu$^{2+}$ concentrations using O and/or Si as co-dopants and suitable experimental conditions. Oxygen was found to play a critical role in the incorporation of Eu into GaN and the quality of Eu-doped GaN samples prepared by organo-metallic vapor phase epitaxy (OMVPE)~\cite{Mitchell2016SR,Mitchell2017MCP} and lead to sharp and uniform emission spectra and improved energy transfer efficiency \cite{Mitchell2016SR,Zhu2016APLM,Mitchell2017MCP}. Significant enhancement of the luminescence intensity were also found in GaN co-doped with Eu and Mg~\cite{Takagi2011APL,Lee2012APL,Wakahara2012JL,Sekiguchi2013JAP,Sekiguchi2016APL} or Si~\cite{Wang2009JAP}. Notably, Mg-containing Eu-doped GaN samples were reported to exhibit photoluminescence (PL) hysteresis through the so-called ``hysteretic photochromic switching'' according to which the observed temperature dependence of PL during a cooling-warming cycle was thought to be due to a switching between two different configurations of an Eu-Mg defect~\cite{ODonnell2014PSSC,ODonnell2016APL,Singh2017SC,Cameron2020APL}. 

Altogether the luminescence in Eu-doped GaN can be characterized by its complexity with the presence of multiple optically active centers and the dependence on the growth conditions. The interpretation of experimental observations and the discussion in terms of specific defect configurations have been, however, largely speculative.

On the theory side, calculations for Eu-doped GaN were carried out by several research groups using density-functional theory (DFT) based methods, including the local-density approximation (LDA) or self-interaction corrected LDA, the generalized gradient approximation (GGA), and GGA$+U$, and LDA$+U$ within a DFT-based tight-binding approach~\cite{Filhol2004,Svane2006,Sanna2009,Ouma2014PhysicaB,Mitchell2017PRB,GoumriSaid2008JPD,Su2018PhysicaB,Cruz2012PRB,Masago2014JJAP,Masago2014APEx,Masago2014APE,Masago2017APE}. These studies provided some information on the structural and electronic properties and limited data on defect structure and energetics. Besides, the methods employed in these studies are known to have limited predictive power, especially in determining defect energy levels \cite{Freysoldt2014RMP}; see the Supplemental Material (SM) \cite{SM} for a detailed discussion. A more rigorous theoretical and computational approach is needed for the study of defect physics in RE-doped semiconductors. In such an approach, the employed methods should possess the ability to overcome the ``band-gap problem'' encountered in DFT (within LDA or GGA) and DFT+$U$ calculations and, at the same time, provide a good description of the structural and electronic properties of the RE-doped systems, including local defect structure.

Here, we present a first-principles investigation of defect physics in Eu-doped GaN using hybrid density-functional defect calculations. In these calculations, all orbitals in the material are treated on equal footing, unlike in DFT$+$$U$ calculations where the Hubbard $U$ term is applied on the RE $4f$ states only and all other orbitals are left uncorrected. The hybrid DFT/Hartree-Fock method \cite{heyd:8207} employed here has been shown to be superior to DFT and DFT$+$$U$ in the study of defects in semiconductors in general~\cite{Freysoldt2014RMP} and RE-doped materials in particular~\cite{Hoang2016RRL}. Specific calculations are carried out for the substitutional Eu impurity, native defects (Ga and N vacancies), and impurities (O, Si, C, H, and Mg) in both the unassociated (i.e., isolated defect) form and the associated (i.e., defect complex) form. These impurities are selected as they are common unintentional or intentional co-dopants in GaN. Based on the results, we discuss the tuning of the valence state and concentration of Eu through co-doping and defect association and examine the role of Eu-related defects as carrier traps for intra-$f$ luminescence and defect-to-band transitions. 

\section{Methodology}\label{sec;method} 

We model defects in the GaN host using a supercell approach in which a defect is included in a periodically repeated finite volume of the host material. Note that we often use ``defect'' as a generic term, referring to not only native point defects (intrinsic to the materials) but also impurities (i.e., extrinsic point defects), and defect complexes; impurities can be intentionally incorporated (i.e., dopants) or unintentionally present. The formation energy of a defect X in effective charge state $q$ (i.e., with respect to the host lattice) is defined as \cite{walle:3851,Freysoldt2014RMP}     
\begin{align}\label{eq:eform}
E^f({\mathrm{X}}^q)&=&E_{\mathrm{tot}}({\mathrm{X}}^q)-E_{\mathrm{tot}}({\mathrm{bulk}}) -\sum_{i}{n_i\mu_i} \\ %
\nonumber &&+~q(E_{\mathrm{v}}+\mu_{e})+ \Delta^q ,
\end{align}
where $E_{\mathrm{tot}}(\mathrm{X}^{q})$ and $E_{\mathrm{tot}}(\mathrm{bulk})$ are the total energies of the defect and bulk supercells; $n_{i}$ is the number of atoms of species $i$ that have been added ($n_{i}>0$) or removed ($n_{i}<0$) to form the defect; $\mu_{i}$ is the atomic chemical potential, representing the energy of the reservoir with which atoms are being exchanged. $\mu_{e}$ is the electronic chemical potential, i.e., the Fermi level, representing the energy of the electron reservoir, referenced to the valence-band maximum (VBM) in the bulk ($E_{\mathrm{v}}$). Finally, $\Delta^q$ is the correction term to align the electrostatic potentials of the bulk and defect supercells and to account for finite-size effects on the total energies of charged defects, calculated following the procedure of Freysodt et al.~\cite{Freysoldt,Freysoldt11}.

In thermodynamic {\it equilibrium}, the formation energy of a defect directly determines the concentration \cite{walle:3851}:
\begin{equation}\label{eq;con} 
c=N_{\rm sites}N_{\rm config}\exp{\left(\frac{-E^f}{k_{\rm B}T}\right)}, 
\end{equation} 
where $N_{\rm sites}$ is the number of high-symmetry sites in the lattice (per unit volume) on which the defect can be incorporated, $N_{\rm config}$ is the number of equivalent configurations (per site), and $k_{\rm B}$ is the Boltzmann constant. Clearly, at a given temperature, a defect that has a lower formation energy will be more likely to form and occur with a higher concentration. Note that, when a material is prepared under {\it non-equilibrium} conditions, excess defects can be frozen-in and the equilibrium concentration estimated via Eq.~(\ref{eq;con}) is only the {\it lower bound} \cite{Hoang2018JPCM}. 

While the Fermi level in Eq.~(\ref{eq:eform}) can be treated as a variable, it is not a free parameter. The actual Fermi-level position can be determined by solving the charge-neutrality equation \cite{walle:3851}:
\begin{equation}\label{eq:neutrality}
\sum_{i}c_{i}q_{i}-n_{e}+n_{h}=0,
\end{equation}
where $c_{i}$ and $q_{i}$ are the concentration and charge, respectively, of defect X$_{i}$; $n_{e}$ and $n_{h}$ are free electron and hole concentrations, respectively; and the summation is over all possible defects present in the material

From defect formation energies, one can calculate the {\it thermodynamic} transition level between charge states $q$ and $q'$ of a defect, $\epsilon(q/q')$, defined as the Fermi-level position at which the formation energy of the defect in charge state $q$ is equal to that in charge state $q'$ \cite{Freysoldt2014RMP}, i.e.,
\begin{equation}\label{eq;tl}
\epsilon(q/q') = \frac{E^f(X^{q}; \mu_e=0)-E^f(X^{q'}; \mu_e=0)}{q' - q},
\end{equation}
where $E^f(X^{q}; \mu_e=0)$ is the formation energy of the defect X in charge state $q$ when the Fermi level is at the VBM ($\mu_e=0$). This $\epsilon(q/q')$ level [often referred to as the $(q/q')$ level], corresponding to a {\it defect energy level} (or simply {\it defect level}), would be observed in, e.g., deep-level transient spectroscopy (DLTS) experiments where the defect in the final charge state $q'$ fully relaxes to its equilibrium configuration after the transition. Note that these defect levels are {\it not} the same as the Kohn-Sham levels obtained in a band-structure calculation such as those associated with the so-called ``defect states'' that may be observed in the electronic density of states (DOS) of a system in the presence of a defect. Strictly speaking, the Kohn-Sham levels cannot be directly identified with any levels that can be observed in experiments \cite{walle:3851,Freysoldt2014RMP}.

The {\it optical} transition level $E_{\rm opt}^{q/q'}$ is defined similarly but with the total energy of the final state $q'$ calculated using the lattice configuration of the initial state $q$ \cite{Freysoldt2014RMP}.     

Our total-energy calculations are based on DFT with the Heyd-Scuseria-Ernzerhof (HSE) functional \cite{heyd:8207}, the projector augmented wave method \cite{PAW2}, and a plane-wave basis set, as implemented in the Vienna {\it Ab Initio} Simulation Package (\textsc{vasp}) \cite{VASP2}. Along with the CPU version, the graphics processing unit (GPU) port \cite{Hacene2012JCC,Hutchinson2012CPC} of \textsc{vasp} is also used. The Hartree-Fock mixing parameter is set to 0.31 and the screening length to the default value of 10 {\AA}. These parameters result in a band gap of 3.53 eV for GaN, very close to that ($\sim$3.5 eV) reported in experiments. Defects in GaN are simulated using a 96-atom supercell and a 2$\times$2$\times$2 Monkhorst-Pack $k$-point mesh for the integrations over the Brillouin zone. In defect calculations, the lattice parameters are fixed to the calculated bulk values but all the internal coordinates are relaxed. The Eu $4f$ electrons are included explicitly in the calculations since in Eu-doped GaN the $4f$ states are present in the band gap and play an important role in defect formation (This is different from, e.g., the case of erbium (Er) doped GaN in which Er $4f$ electrons can be included in the core \cite{Hoang2015RRL}). The Ga $3d$ electrons are treated as core states as the inclusion of these electrons in the valence has small effects on the defect transition level; see the SM \cite{SM} for more details. In all calculations, the plane-wave basis-set cutoff is set to 400 eV and spin polarization is included. All structural relaxations are performed with HSE and the force threshold is chosen to be 0.04 eV/{\AA} or smaller. Spin-orbit coupling (SOC) is not included since significant cancellation is expected between the terms in Eqs.~(\ref{eq:eform}) and (\ref{eq;tl}). Our tests show that the $\epsilon(0/-)$ level of Eu$_{\rm Ga}$ obtained in HSE$+$SOC calculations is different from that obtained in HSE calculations by only 7 meV.    

The chemical potentials of Ga, N, Eu, H, C, O, Si, and Mg are referenced to the total energy per atom of bulk Ga, N$_2$ at 0 K, bulk Eu, H$_2$ at 0 K, bulk C (diamond), O$_2$ at 0 K, bulk Si, and bulk Mg, respectively. $\mu_{\rm Ga}$ and $\mu_{\rm N}$ vary over a range determined by the formation enthalpy of GaN such that $\mu_{\rm Ga}+\mu_{\rm N} = \Delta H({\rm GaN})$ (calculated to be $-$1.26 eV at 0 K). We will examine defect landscape in GaN in two extreme limits: Ga-rich ($\mu_{\rm Ga} = 0$) and N-rich ($\mu_{\rm N} = 0 $) conditions. Specific values of the Eu, H, C, O, Si, and Mg chemical potentials are determined by assuming equilibrium with EuN ($\Delta H = -1.20$ eV at 0 K), H$_2$ at 0 K, bulk C, $\beta$-Ga$_2$O$_{3}$ ($-$10.07 eV), $\beta$-Si$_3$N$_4$ ($-$9.17 eV), and Mg$_3$N$_2$ ($-$4.16 eV), respectively. Note that the transition levels $\epsilon(q/q')$ and $E_{\rm opt}^{q/q'}$ are {\it independent} of the choice of the atomic chemical potentials. %$\mu_i$. 

\section{Results}\label{sec;results}

\subsection{Unassociated native defects and impurities}\label{sec;isolated}

In bulk (i.e., perfect and undoped) GaN (wurtzite, space group $P6_3mc$), each Ga is coordinated with four N atoms: one along the $c$-axis and three in the basal ($ab$) plane. There is a small $C_{3v}$ distortion at the Ga lattice site with the Ga$-$N bond length calculated for the axial N atom (1.958 {\AA}) slightly different from that for the basal N atoms (1.952 {\AA}). For comparison, the experimental values for the axial and basal Ga$-$N bond lengths are 1.956 {\AA} and 1.949 {\AA} \cite{Schulz1977SSC}, respectively. In the presence of a defect, the lattice environment in the defect's vicinity can be further distorted, and such a local distortion is often different for different charge configurations. 

\begin{figure}[t]%
\vspace{0.2cm}
\includegraphics*[width=\linewidth]{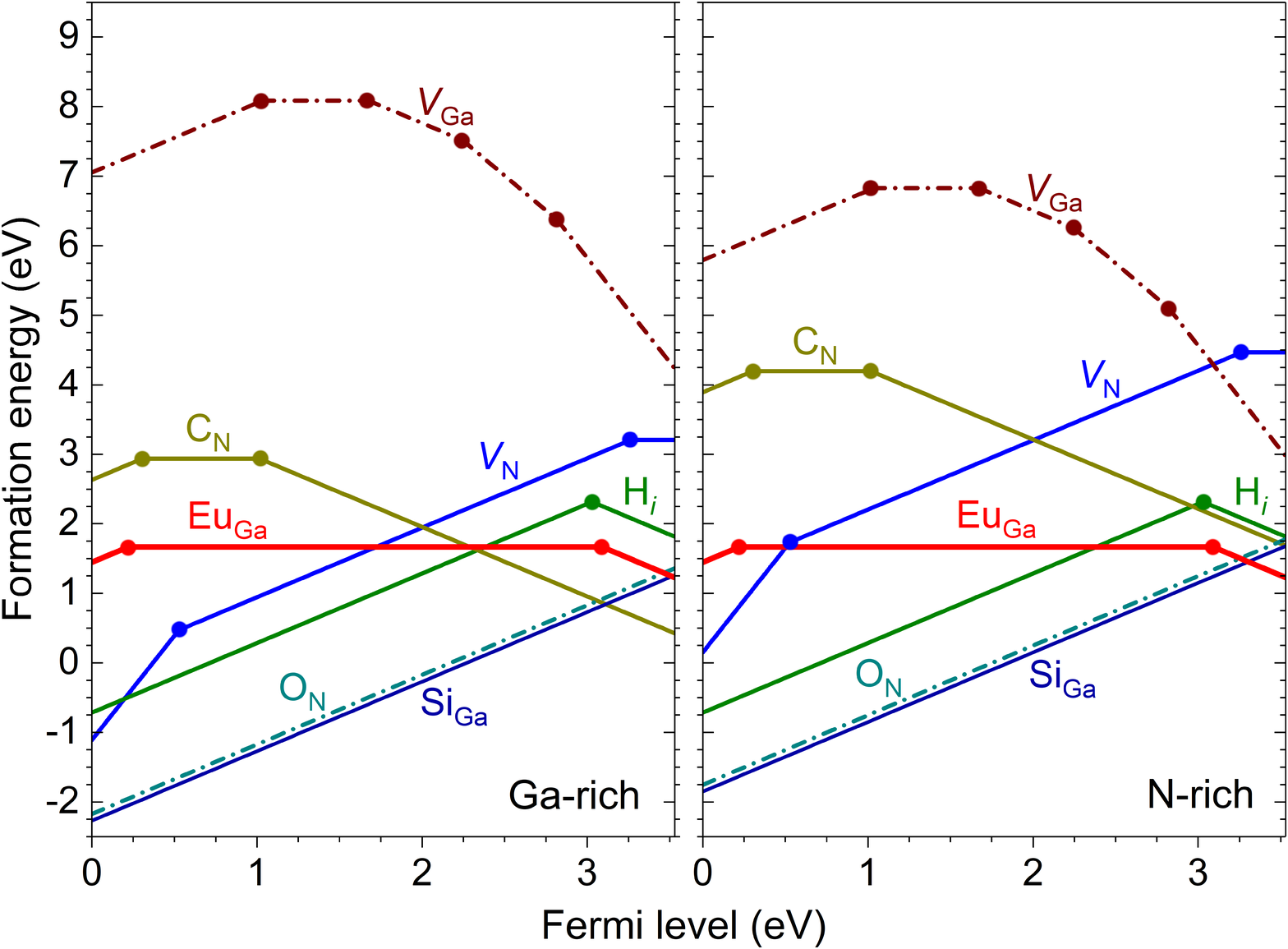}
\caption{Formation energies of Eu$_{\rm Ga}$ and relevant native point defects ($V_{\rm N}$, $V_{\rm Ga}$) and impurities (H$_i$, C$_{\rm N}$, O$_{\rm N}$, Si$_{\rm Ga}$) in GaN, plotted as a function of Fermi level from the VBM (at 0 eV) to the conduction-band minimum (CBM, at 3.53 eV), under the extreme Ga-rich and N-rich conditions. For each defect, only segments corresponding to the lowest-energy charge states are shown. The slope of these segments indicates the charge state [i.e., $q$ in Eq.~(\ref{eq:eform})]: positively (negatively) charged defect configurations have positive (negative) slopes; horizontal segments correspond to neutral defect conﬁgurations. Large solid dots connecting two segments with different slopes, if present, mark the {\it defect levels} [i.e., the thermodynamic transition levels, $\epsilon(q/q')$, calculated according to Eq.~(\ref{eq;tl})].}
\label{fig;fe1} 
\end{figure}

Figure~\ref{fig;fe1} shows the formation energies of various {\it unassociated} native defects and impurities in GaN. The substitutional Eu impurity (Eu$_{\rm Ga}$) is found to be stable as Eu$_{\rm Ga}^0$ (i.e., Eu$^{3+}$ at the Ga$^{3+}$ lattice site, with a calculated magnetic moment of 6 $\mu_{\rm B}$; spin $S=3$) and/or Eu$_{\rm Ga}^-$ (i.e., Eu$^{2+}$ at the Ga$^{3+}$ site, with a magnetic moment of 7 $\mu_{\rm B}$; spin $S=7/2$), depending on the Fermi-level position. Eu$_{\rm Ga}$ introduces two defect levels in the host band gap: the $(+/0)$ level at 0.22 eV above the VBM and the $(0/-)$ level at 3.09 eV (i.e., 0.44 eV below the CBM). Note, however, that ``Eu$_{\rm Ga}^+$'' is not a true charge state of Eu$_{\rm Ga}$. It is, in fact, a defect complex consisting of Eu$_{\rm Ga}^0$ and an electron hole ($h^\ast$; spin $S=1/2$) {\it localized} on one of the neighboring basal N atoms. In the Eu$_{\rm Ga}^0$ (Eu$_{\rm Ga}^-$) configuration, the Eu$-$N bond length is 2.233 {\AA} (2.321 {\AA}) for the axial N atom and 2.197$-$2.208 {\AA} (2.283$-$2.297 {\AA}) for the basal N atoms. The local distortion at the Ga lattice site where the Eu dopant is incorporated is thus more pronounced and slightly deviates from the $C_{3v}$ symmetry. The result is also consistent with the fact that the ionic radius of Eu$^{3+}$ is smaller than that of Eu$^{2+}$. 

The electronic behavior of Eu is thus different from that of, e.g., erbium (Er) in GaN. Er was found to be stable only as Er$^{3+}$, and the unassociated Er$_{\rm Ga}$ does not introduce any defect levels \cite{Hoang2015RRL}. The origin of the difference can be traced back to the difference in their electronic structure: Eu introduces an unoccupied $4f$ state in the host band gap, whereas Er does not produce any in-gap $4f$ state; see the SM \cite{SM} for a detailed analysis of the electronic structure of Eu- {\it vs}.~Er-doped GaN obtained in DFT$+$$U$ and HSE calculations. As a result, upon adding an electron to Eu$_{\rm Ga}^0$ to form Eu$_{\rm Ga}^-$ the extra electron goes to the lowest unoccupied state (in this case, the in-gap Eu $4f$ state), leading to the valence change from Eu$^{3+}$ to Eu$^{2+}$; see Fig.~4 of the SM \cite{SM}. In the case of Er, the extra electron goes to the CBM (composed of the host states) and becomes delocalized; the valence change, therefore, does not occur and ``Er$_{\rm Ga}^-$'' cannot be stabilized \cite{Hoang2015RRL}. The implications of the $(0/-)$ level of Eu$_{\rm Ga}$ and the valence change are discussed in Sec.~\ref{sec;dis}.

Our result for Eu$_{\rm Ga}$ is {\it significantly} different from that previously reported in the literature \cite{Filhol2004,Svane2006,Sanna2009}. Filhol et al.~\cite{Filhol2004} and Svane et al.~\cite{Svane2006} did not find any defect energy level in the calculated band gap, i.e., Eu is stable as Eu$^{3+}$ in the entire range of the Fermi-level values from the VBM to the CBM, which is in contrast to the fact that Eu$^{2+}$ is also stable in GaN. Sanna et al.~\cite{Sanna2009}, on the other hand, reported the $(0/-)$ level of Eu$_{\rm Ga}$ to be at 1.58 eV above the VBM, which is about 1.5 eV lower than that found in our calculations. A more detailed discussion of those previous studies is provided in the SM \cite{SM}.        

The results for the nitrogen vacancy ($V_{\rm N}$), gallium vacancy ($V_{\rm Ga}$), hydrogen interstital (H$_i$), and substitutional carbon (C$_{\rm N}$) and oxygen (O$_{\rm N}$) were already reported and discussed in detail in Ref.~\cite{Hoang2015RRL} but are included in Fig.~\ref{fig;fe1} for easy reference since in the next sections we will discuss defect complexes consisting of Eu$_{\rm Ga}$ and these native point defects and impurities. $V_{\rm N}$ introduces the $(3+/+)$ level at 0.53 eV above the VBM and the $(+/0)$ level at 0.27 eV below the CBM. $V_{\rm Ga}$ has four defect levels in the band gap: $(+/0)$ at 1.03 eV and $(0/-)$ at 1.67 eV above the VBM, and $(-/2-)$ at 1.29 eV and $(2-/3-)$ at 0.71 eV below the CBM. H$_i$ is amphoteric [i.e., positively (negatively) charged in the $p$-type ($n$-type) GaN] and its $(+/-)$ level occurs at 0.50 eV below the CBM. C$_{\rm N}$ has two defect levels: $(+/0)$ at 0.31 eV and $(0/-)$ at 1.02 eV above the VBM. O$_{\rm N}$ is a shallow donor and only stable as O$_{\rm N}^+$ \cite{Hoang2015RRL}. We find that Si$_{\rm Ga}$ is also a shallow donor, being stable only in the Si$_{\rm Ga}^+$ configuration. Si (O) thus readily donates one electron to the lattice and becomes a positively charged defect when incorporated at the Ga (N) lattice site in GaN. The result for Si$_{\rm Ga}$ is in agreement with that previously reported by other groups \cite{Park1997PRB,VandeWalle1998PRB,Gordon2014PRB}. 

\subsection{Defect complexes of Eu and native defects}\label{sec;withnative}

\begin{figure}[t]%
\vspace{0.2cm}
\includegraphics*[width=\linewidth]{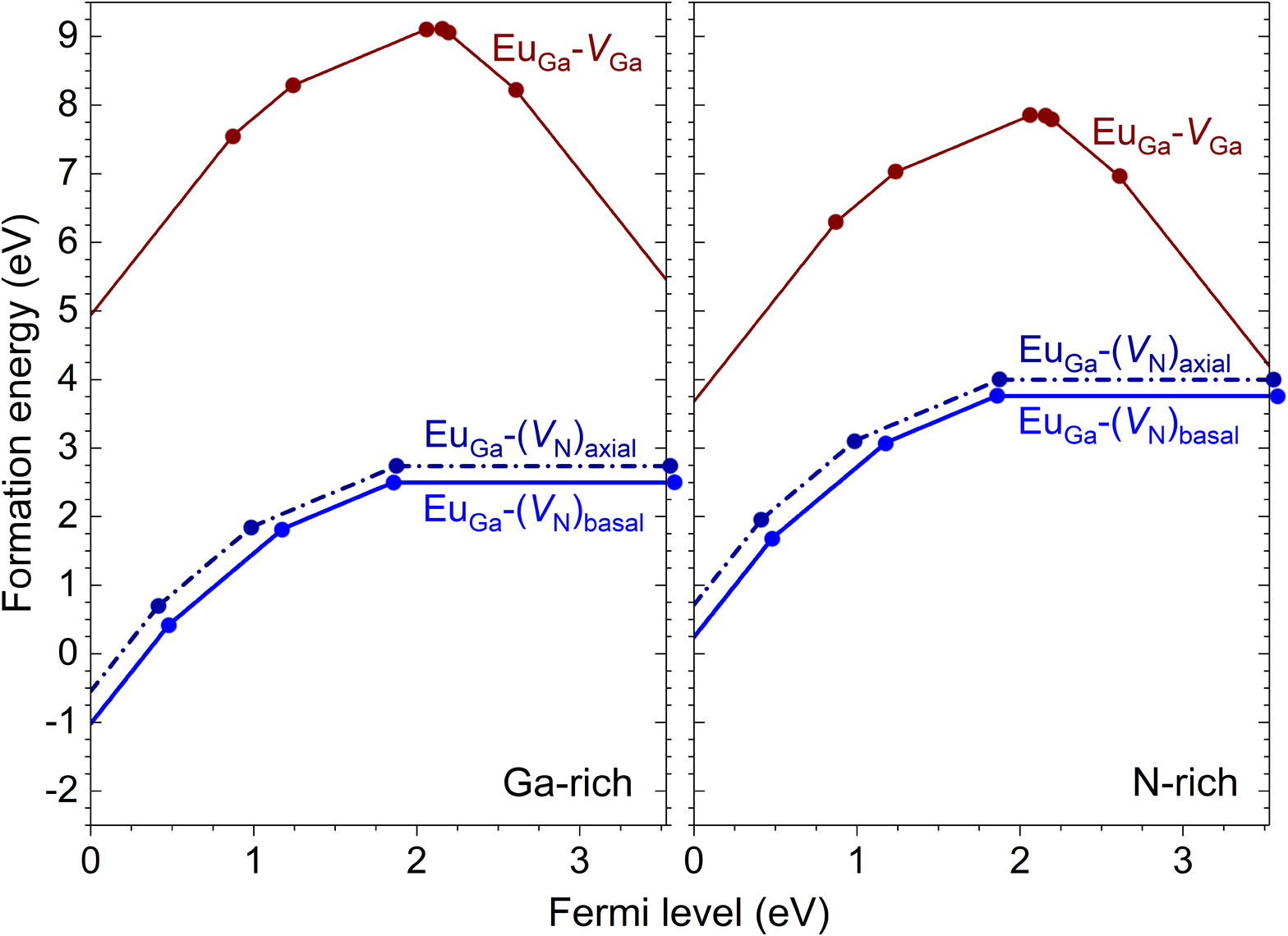}
\caption{Formation energies of defect complexes consisting of Eu$_{\rm Ga}$ and native defects in GaN, Eu$_{\rm Ga}$-$V_{\rm N}$ and Eu$_{\rm Ga}$-$V_{\rm Ga}$, plotted as a function of Fermi level from the VBM to the CBM, under the Ga-rich and N-rich conditions. For each defect, only segments corresponding to the lowest-energy charge states are shown. Large solid dots connecting two segments with different slopes mark the defect levels. $V_{\rm N}$ can be at the basal or axial N site with respect to the Eu$_{\rm Ga}$ component. The $(0/-)$ level of Eu$_{\rm Ga}$-$V_{\rm N}$ is right above the CBM.}
\label{fig;fe2} 
\end{figure}

\begin{figure}[t]%
\vspace{0.2cm}
\includegraphics*[width=\linewidth]{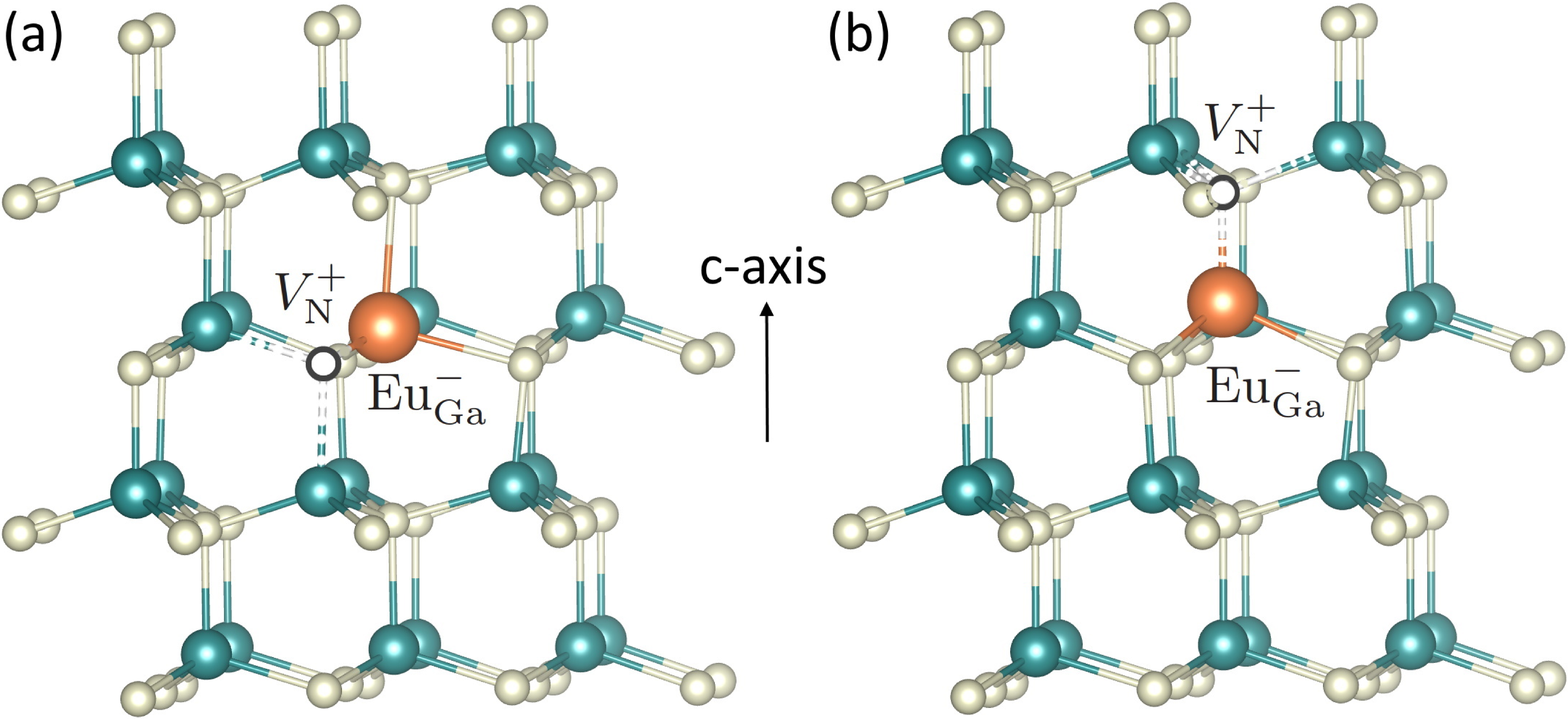}
\caption{Structure of (Eu$_{\rm Ga}$-$V_{\rm N}$)$^0$: (a) basal and (b) axial geometric configurations. Large spheres are Eu, medium Ga, small N. The nitrogen vacancy is represented by a black circle.}
\label{fig;struct1}
\end{figure}

\begin{figure}[t]%
\vspace{0.2cm}
\includegraphics*[width=0.97\linewidth]{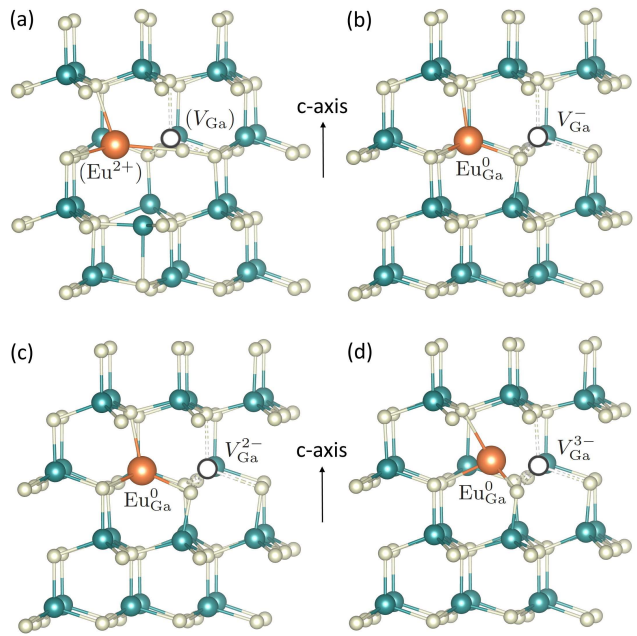}
\caption{Structure of representative (Eu$_{\rm Ga}$-$V_{\rm Ga}$)$^q$ defect configurations: (a) $q=0$, (b) $q=-$, (c) $q=2-$, and (d) $q=3-$. Large spheres are Eu, medium Ga, and small N. The gallium vacancy is represented by a black circle. The structures associated with $q=+,2+,3+$ (not shown here) are similar to that of $q=0$. In (a), the charge state of the $V_{\rm Ga}$ constituent cannot be clearly determined at this point; see the text.}
\label{fig;struct2}
\end{figure}

\begin{table*}%[t!]
\caption{Eu-related defects in Eu-doped GaN: The stable valence state of the rare-earth (RE) ion, constituent defects, binding energy ($E_b$, with respect to the isolated constituents), magnetic moment ($M$), and defect levels [$\epsilon(q/q')$, with respect to the VBM ($E_{\rm v}$, at 0 eV) or the CBM ($E_{\rm c}$, at 3.53 eV)]. Note that $h^\ast$ is an electron hole localized at an N lattice site. Spin-polarized defects in a complex are found to interact ferromagnetically; the magnetic moment of the complex is thus equal to the sum of those of the constituents. The values in the parentheses are for the axial (with respect to Eu) configurations.}\label{tab;complex}
\begin{center}
%\small
\begin{ruledtabular}
%\resizebox{\columnwidth}{!}{%
\begin{tabular}{lllccl}
%\colrule \colrule
Defect &RE ion &Constituents & $E_{b}$ (eV) & $M$ ($\mu_{\rm B}$) & Defect levels (eV) \\
\colrule
Eu$_{\rm Ga}^+$ &Eu$^{3+}$& Eu$_{\rm Ga}^0$ + $h^\ast$ &&7&$\epsilon(+/0)=E_{\rm v}+0.22$\\
Eu$_{\rm Ga}^0$ &Eu$^{3+}$& Eu$_{\rm Ga}^0$ && 6&$\epsilon(0/-)=E_{\rm c}-0.44$\\
Eu$_{\rm Ga}^-$ &Eu$^{2+}$& Eu$_{\rm Ga}^-$ &&7&\\ 
\\
(Eu$_{\rm Ga}$-$V_{\rm N}$)$^{3+}$ &Eu$^{3+}$& Eu$_{\rm Ga}^0$ + $V_{\rm N}^{3+}$ &1.58 (1.11)&6&$\epsilon(3+/2+)=E_{\rm v}+0.48$ $(0.42)$ \\
(Eu$_{\rm Ga}$-$V_{\rm N}$)$^{2+}$ &Eu$^{2+ (3+)}$& Eu$_{\rm Ga}^{- (0)}$ + $V_{\rm N}^{3+ (2+)}$ &4.18 (1.25)&7&$\epsilon(2+/+)=E_{\rm v}+1.18$ $(0.99)$ \\
(Eu$_{\rm Ga}$-$V_{\rm N}$)$^+$ &Eu$^{3+}$& Eu$_{\rm Ga}^0$ + $V_{\rm N}^+$ &0.97 (0.75)&6&$\epsilon(+/0)=E_{\rm c}-1.67$ $(1.65)$\\
(Eu$_{\rm Ga}$-$V_{\rm N}$)$^0$ &Eu$^{2+}$& Eu$_{\rm Ga}^-$ + $V_{\rm N}^+$ &2.20 (1.96)&7&$\epsilon(0/-)=E_{\rm c}+0.07$ $(0.03)$\\
(Eu$_{\rm Ga}$-$V_{\rm N}$)$^-$ &Eu$^{2+}$& Eu$_{\rm Ga}^-$ + $V_{\rm N}^0$ &1.86 (1.67)&8&\\
\\
(Eu$_{\rm Ga}$-$V_{\rm Ga}$)$^{3+}$ &Eu$^{3+}$&&&6&$\epsilon(3+/2+)=E_{\rm v}+0.87$\\ 
(Eu$_{\rm Ga}$-$V_{\rm Ga}$)$^{2+}$ &Eu$^{3+}$&&&7&$\epsilon(2+/+)=E_{\rm v}+1.23$\\ 
(Eu$_{\rm Ga}$-$V_{\rm Ga}$)$^+$ &Eu$^{2+}$&&&8&$\epsilon(+/0)=E_{\rm v}+2.06$\\ 
(Eu$_{\rm Ga}$-$V_{\rm Ga}$)$^0$ &Eu$^{2+}$&&&9&$\epsilon(0/-)=E_{\rm c}-1.38$\\
(Eu$_{\rm Ga}$-$V_{\rm Ga}$)$^-$ &Eu$^{3+}$& Eu$_{\rm Ga}^0$ + $V_{\rm Ga}^-$ &0.18&8&$\epsilon(-/2-)=E_{\rm c}-1.34$\\
(Eu$_{\rm Ga}$-$V_{\rm Ga}$)$^{2-}$ &Eu$^{3+}$& Eu$_{\rm Ga}^0$ + $V_{\rm Ga}^{2-}$ &0.23&7&$\epsilon(2-/3-)=E_{\rm c}-0.92$\\
(Eu$_{\rm Ga}$-$V_{\rm Ga}$)$^{3-}$ &Eu$^{3+}$& Eu$_{\rm Ga}^0$ + $V_{\rm Ga}^{3-}$ &0.45&6&\\
\\
(Eu$_{\rm Ga}$-H$_{i}$)$^+$ &Eu$^{3+}$& Eu$_{\rm Ga}^0$ + H$_{i}^+$ &1.25&6&$\epsilon(+/0)=E_{\rm c}-1.36$\\
(Eu$_{\rm Ga}$-H$_{i}$)$^0$ &Eu$^{2+}$& Eu$_{\rm Ga}^-$ + H$_{i}^+$ &2.17&7&\\
\\
(Eu$_{\rm Ga}$-C$_{\rm N}$)$^+$ &Eu$^{3+}$& Eu$_{\rm Ga}^0$ + C$_{\rm N}^+$ &0.61 (0.50)&8&$\epsilon(+/0)=E_{\rm v}+0.33$ $(0.17)$\\
(Eu$_{\rm Ga}$-C$_{\rm N}$)$^0$ &Eu$^{3+}$& Eu$_{\rm Ga}^0$ + C$_{\rm N}^0$ &0.58 (0.63)&7&$\epsilon(0/-)=E_{\rm c}-1.08$ $(1.17)$\\
(Eu$_{\rm Ga}$-C$_{\rm N}$)$^-$ &Eu$^{3+}$& Eu$_{\rm Ga}^0$ + C$_{\rm N}^-$ &$-$0.86 ($-$0.71)&6&\\
\\
(Eu$_{\rm Ga}$-O$_{\rm N}$)$^+$ &Eu$^{3+}$& Eu$_{\rm Ga}^0$ + O$_{\rm N}^+$ &0.76 (0.62)&6&$\epsilon(+/0)=E_{\rm c}-1.26$ $(1.23)$\\
(Eu$_{\rm Ga}$-O$_{\rm N}$)$^0$ &Eu$^{2+}$& Eu$_{\rm Ga}^-$ + O$_{\rm N}^+$ &1.59 (1.41)&7&\\
\\
(Eu$_{\rm Ga}$-Si$_{\rm Ga}$)$^+$ &Eu$^{3+}$& Eu$_{\rm Ga}^0$ + Si$_{\rm Ga}^+$ &0.32&6&$\epsilon(+/0)=E_{\rm c}-1.04$\\
(Eu$_{\rm Ga}$-Si$_{\rm Ga}$)$^0$ &Eu$^{2+}$& Eu$_{\rm Ga}^-$ + Si$_{\rm Ga}^+$ &0.92&7&\\
\\
(Eu$_{\rm Ga}$-Mg$_{\rm Ga}$)$^0$ &Eu$^{3+}$& Eu$_{\rm Ga}^0$ + Mg$_{\rm Ga}^-$ + $h^\ast$ &0.48 (0.40)$^a$&7&$\epsilon(0/-)=E_{\rm v}+0.97$ $(0.85)$\\
(Eu$_{\rm Ga}$-Mg$_{\rm Ga}$)$^-$ &Eu$^{3+}$& Eu$_{\rm Ga}^0$ + Mg$_{\rm Ga}^-$ &$-$0.10 ($-$0.06)&6&\\
\\
(Eu$_{\rm Ga}$-O$_{\rm N}$-Mg$_{\rm Ga}$)$^0$ &Eu$^{3+}$& Eu$_{\rm Ga}^0$ + O$_{\rm N}^+$ + Mg$_{\rm Ga}^-$ &1.70&6&$\epsilon(0/-)=E_{\rm c}-0.78$\\
(Eu$_{\rm Ga}$-O$_{\rm N}$-Mg$_{\rm Ga}$)$^-$ &Eu$^{2+}$& Eu$_{\rm Ga}^-$ + O$_{\rm N}^+$ + Mg$_{\rm Ga}^-$ &2.04&7&\\
%\colrule \colrule
\end{tabular}
%}
\end{ruledtabular}
\end{center}
\begin{flushleft}
$^a$With respect to Eu$_{\rm Ga}^0$ and ``Mg$_{\rm Ga}^0$''.
\end{flushleft}
\end{table*}

Some of the unassociated defects discussed in Sec.~\ref{sec;isolated} can come close and form complexes. Such defect association often changes the local lattice environment and defect energetics and can lead to important implications. Among possible complexes between Eu$_{\rm Ga}$ and native defects, Eu$_{\rm Ga}$-$V_{\rm N}$ and Eu$_{\rm Ga}$-$V_{\rm Ga}$ have been widely thought to be possible defect centers for intra-$f$ luminescence in Eu-doped GaN \cite{Roqan2010PRB,Fujiwara2014JJAP,Mitchell2014JAP,Mitchell2017PRB,Mitchell2018JAP}. Figure \ref{fig;fe2} shows the formation energies of these complexes. In Eu$_{\rm Ga}$-$V_{\rm N}$, the $V_{\rm N}$ part can be at the basal or axial lattice site with respect to Eu$_{\rm Ga}$; see Fig.~\ref{fig;struct1}. We find that Eu$_{\rm Ga}$-($V_{\rm N}$)$_{\rm basal}$ introduces four defect levels in the band gap region: $(3+/2+)$ at 0.48 eV, $(2+/+)$ at 1.18 eV, and $(+/0)$ at 1.86 eV above the VBM, and $(0/-)$ at 0.07 eV above the CBM. A careful inspection shows that in going from (Eu$_{\rm Ga}$-$V_{\rm N}$)$^0$ to (Eu$_{\rm Ga}$-$V_{\rm N}$)$^-$ the additional electron stays in the vicinity of the void formed by $V_{\rm N}$; (Eu$_{\rm Ga}$-$V_{\rm N}$)$^0$ is a defect complex consisting of Eu$_{\rm Ga}^-$ and $V_{\rm N}^+$ whereas (Eu$_{\rm Ga}$-$V_{\rm N}$)$^-$ is a complex of Eu$_{\rm Ga}^-$ and $V_{\rm N}^0$; see also Table \ref{tab;complex}. The extra electron is thus captured by the $V_{\rm N}$ part of the complex and Eu at the Ga site remains Eu$^{2+}$. The axial geometric configuration, i.e., Eu$_{\rm Ga}$-($V_{\rm N}$)$_{\rm axial}$, has a higher formation energy than the basal one in all the stable charge states; its defect transition levels are also slightly shifted as seen in Fig.~\ref{fig;fe2}; e.g., the $(0/-)$ level is now at 0.03 eV above the CBM. Eu$_{\rm Ga}$-$V_{\rm Ga}$, on the other hand, has six defect levels in the host band gap: $(3+/2+)$ at 0.87 eV, $(2+/+)$ at 1.23 eV, $(+/0)$ at 2.06 eV, $(0/-)$ at 2.15 eV, $(-/2-)$ at 2.19 eV, and $(2-/3-)$ at 2.61 eV above the VBM. 

The local lattice environment is changed significantly due to defect--defect interaction. In (Eu$_{\rm Ga}$-$V_{\rm N}$)$^0$, for example, Eu  moves off-center and closer to the vacancy by 0.27 {\AA}; see Fig.~\ref{fig;struct1}(a). The binding energy of the complex with respect to its isolated constituents, Eu$_{\rm Ga}^-$ and $V_{\rm N}^+$, is 2.20 eV. In the other stable charge states of the basal configuration, (Eu$_{\rm Ga}$-$V_{\rm N}$)$^q$ with $q = +, 2+,$ and $3+$, the displacement is 0.19 {\AA}, 0.58 {\AA}, and 0.38 {\AA}, respectively. Similar distortion is observed in the axial geometric configurations; e.g., the displacement of Eu in (Eu$_{\rm Ga}$-$V_{\rm N}$)$^0$ is 0.25 {\AA} along the $c$-axis and toward the vacancy; see Fig.~\ref{fig;struct1}(b). Note that there is significant difference between the basal and axial configurations in the case of (Eu$_{\rm Ga}$-$V_{\rm N}$)$^{2+}$ where Eu is stable as Eu$^{2+}$ in the former and as Eu$^{3+}$ in the latter; see more details in the SM \cite{SM}. The local distortion is generally larger in (Eu$_{\rm Ga}$-$V_{\rm Ga}$)$^q$ where Eu moves off-center by 0.80 {\AA} ($q=0$), 0.34 {\AA} ($-$), 0.49 {\AA} ($2-$), and 1.33 {\AA} ($3-$). The position of their neighboring atoms is also shifted; see Figs.~\ref{fig;struct1} and \ref{fig;struct2}.

The structure and energetics of a defect complex, in general, can be expressed in terms of those of its isolated constituents which are usually elementary defects acting as basic building blocks \cite{Wilson2009PRB,Hoang2011CM}. The example involving Eu$_{\rm Ga}$-$V_{\rm N}$ given above is an illustration of such an analysis which is key to understanding complex defect configurations. In Table \ref{tab;complex}, we list the characteristics of all Eu-related defect complex configurations, including the valence state of Eu, constituent defects, and binding energy of the complexes with respect to the isolated constituents. Note that the decomposition of (Eu$_{\rm Ga}$-$V_{\rm Ga}$)$^q$ with $q=0, +, 2+, 3+$ into basic building blocks is not straightforward due to strong lattice distortion, involving not just Eu but also Ga and N atoms, which makes it difficult to clearly identify constituent defects.

The results summarized in Table \ref{tab;complex} show that Eu is stable as Eu$^{3+}$ or Eu$^{2+}$ in Eu$_{\rm Ga}$-$V_{\rm N}$ and Eu$_{\rm Ga}$-$V_{\rm Ga}$, depending on specific charge states. Compared to the unassociated Eu$_{\rm Ga}$, the association between Eu$_{\rm Ga}$ and $V_{\rm N}$ is found to extend the range of Fermi-level values below the CBM in which Eu$^{2+}$ is energetically more stable than Eu$^{3+}$. This is due to the strong Coulomb attraction between Eu$_{\rm Ga}^-$ and $V_{\rm N}^+$ in (Eu$_{\rm Ga}$-$V_{\rm N}$)$^0$, which leads to a larger reduction (due to defect association) in the formation energy of the neutral charge state compared to that of the preceding ($+$) and subsequent ($-$) stable charge states, as reflected in the binding energies reported in Table \ref{tab;complex}. As a result, the ($+/0$) level of Eu$_{\rm Ga}$-$V_{\rm N}$ is shifted toward the VBM and the ($0/-$) level toward the CBM, compared to those of the unassociated Eu$_{\rm Ga}$, thus extending the stability range of (Eu$_{\rm Ga}$-$V_{\rm N}$)$^0$ and hence Eu$^{2+}$; see Fig.~\ref{fig;fe2}. The basal configuration of (Eu$_{\rm Ga}$-$V_{\rm N}$)$^{2+}$, discussed earlier and in the SM \cite{SM}, presents an even more interesting case where Eu$^{2+}$ can be stabilized far from the CBM. In this example, the strong local elastic and electrostatic interactions play a key role in stabilizing Eu$_{\rm Ga}^-$ and $V_{\rm N}^{3+}$ and in lowering the complex's formation energy.

The binding energy of Eu$_{\rm Ga}$-$V_{\rm N}$ is relatively large, suggesting that it can exist as a defect complex in real samples. Eu$_{\rm Ga}$-$V_{\rm Ga}$, on the other hand, has a much smaller binding energy; see Table \ref{tab;complex}. Together with the high calculated formation energy (Fig.~\ref{fig;fe2}), Eu$_{\rm Ga}$-$V_{\rm Ga}$ is unlikely to be stable or occur with a high concentration as a complex under thermodynamic equilibrium growth conditions. It is more likely to be created under non-equilibrium conditions such as during Eu implantation. The same can be said about Eu$_{\rm Ga}$-$V_{\rm N}$ whose formation energy is high under $n$-type conditions; see Fig.~\ref{fig;fe2}. 

The electronic behavior of Eu$_{\rm Ga}$-$V_{\rm N}$ is similar to Er$_{\rm Ga}$-$V_{\rm N}$ in GaN. Er$_{\rm Ga}$-($V_{\rm N}$)$_{\rm basal}$ was reported to also have the $(0/-)$ level at 0.02 eV above the CBM \cite{Hoang2015RRL}. Our results for Eu$_{\rm Ga}$-$V_{\rm N}$ and Eu$_{\rm Ga}$-$V_{\rm Ga}$ are, however, {\it qualitatively} different from those reported by other groups \cite{Filhol2004,Sanna2009,Ouma2014PhysicaB}; see the SM \cite{SM} for a detailed discussion and comparison.

\subsection{Defect complexes of Eu and other impurities}\label{sec;withimpurities}

\begin{figure}[t]%
\vspace{0.2cm}
\includegraphics*[width=\linewidth]{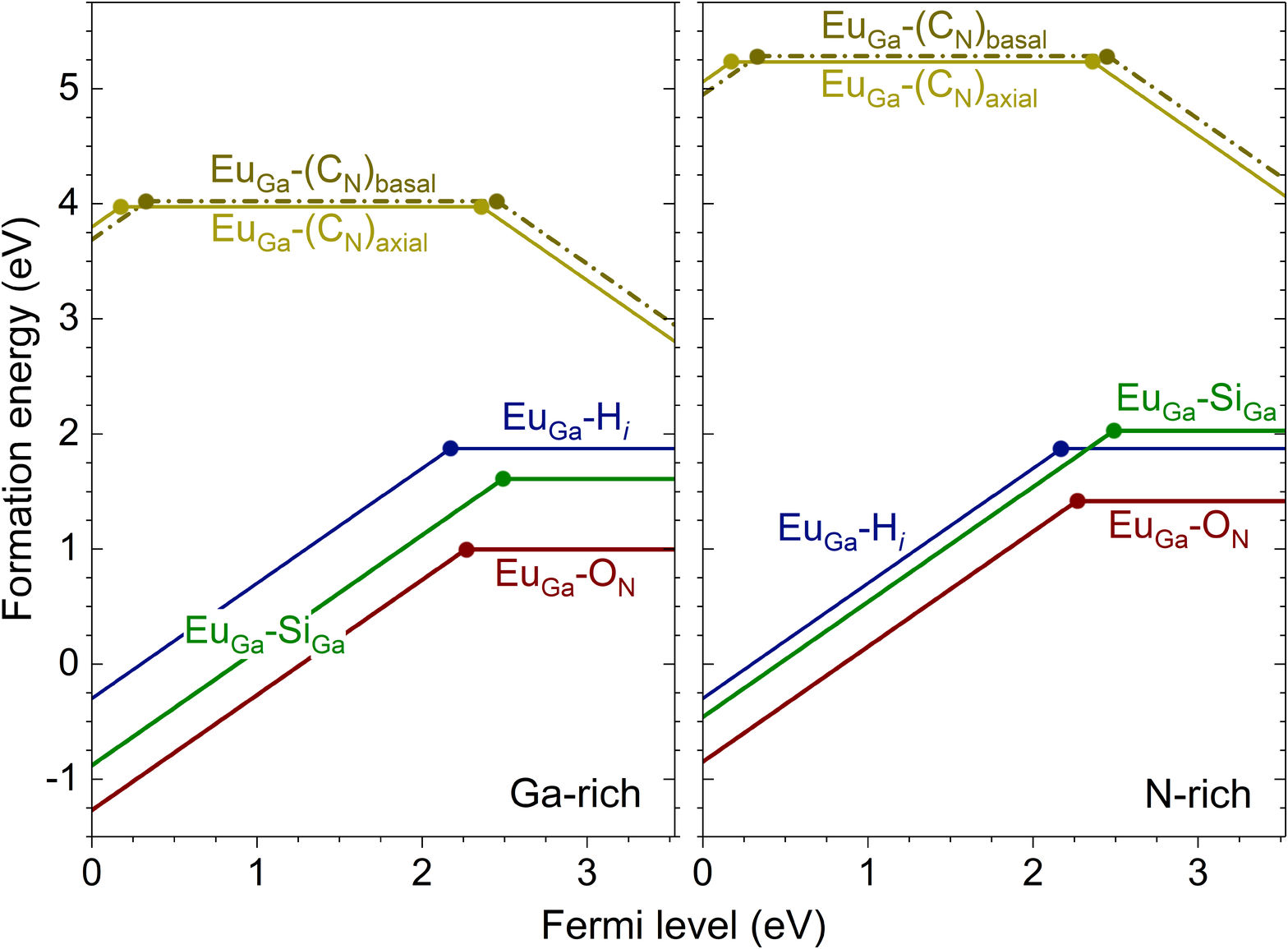}
\caption{Formation energies of defect complexes consisting of Eu$_{\rm Ga}$ and impurities in GaN, Eu$_{\rm Ga}$-H$_i$, Eu$_{\rm Ga}$-C$_{\rm N}$, Eu$_{\rm Ga}$-O$_{\rm N}$, and Eu$_{\rm Ga}$-Si$_{\rm Ga}$, plotted as a function of Fermi level from the VBM to the CBM, under the Ga-rich and N-rich conditions. For each defect, only segments corresponding to the lowest-energy charge states are shown. Large solid dots connecting two segments with diﬀerent slopes mark the defect levels. C$_{\rm N}$ can be at the basal or axial N site with respect to Eu$_{\rm Ga}$. Only the basal configuration if Eu$_{\rm Ga}$-O$_{\rm N}$ is included.}
\label{fig;fe3}
\end{figure}

\begin{figure}[t]%
\vspace{0.2cm}
\includegraphics*[width=\linewidth]{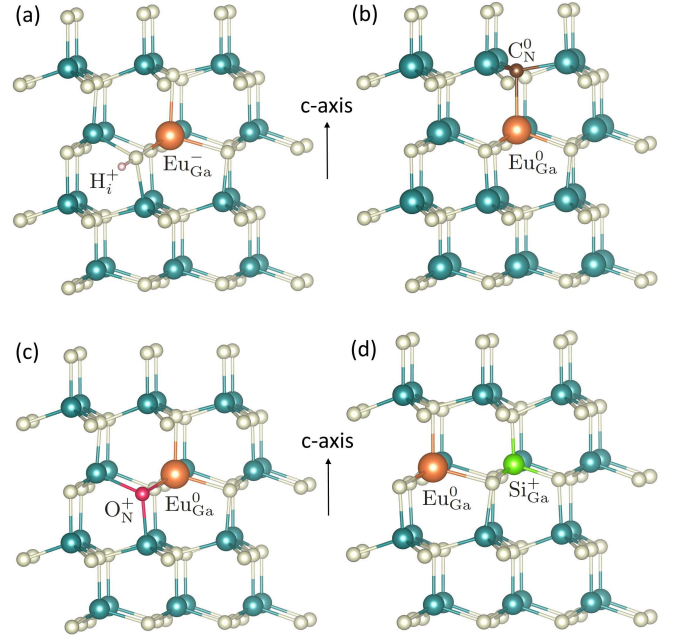}
\caption{Structure of (a) (Eu$_{\rm Ga}$-H$_i$)$^0$, (b) (Eu$_{\rm Ga}$-C$_{\rm N}$)$^0$, (c) (Eu$_{\rm Ga}$-O$_{\rm N}$)$^+$, and (d) (Eu$_{\rm Ga}$-Si$_{\rm Ga}$)$^+$ complexes. Large spheres are Eu, medium Ga/Si, small N/C/O, and smallest H.}
\label{fig;struct3}
\end{figure}

In addition to the native defects, Eu$_{\rm Ga}$ can also form complexes with impurities that are unintentionally present in the growth environment or intentionally incorporated into the host material as co-dopants. Figure \ref{fig;fe3} shows the formation energies of defect complexes Eu$_{\rm Ga}$-H$_i$, Eu$_{\rm Ga}$-C$_{\rm N}$, Eu$_{\rm Ga}$-O$_{\rm N}$, and Eu$_{\rm Ga}$-Si$_{\rm Ga}$. Eu$_{\rm Ga}$-H$_i$ has one defect level in the bulk band gap: $(+/0)$ at 1.36 eV below the CBM. The (Eu$_{\rm Ga}$-H$_i$)$^0$ configuration is a defect complex consisting of Eu$_{\rm Ga}^-$ and H$_i^+$, whereas (Eu$_{\rm Ga}$-H$_i$)$^+$ is a complex of Eu$_{\rm Ga}^0$ and H$_i^+$; see also Table \ref{tab;complex}. The valence state of Eu in the complex thus changes as one crosses the transition level $\epsilon(+/0)$. In these defect complexes, the H interstitial is bonded to one of the nearest N neighbors of Eu. In (Eu$_{\rm Ga}$-H$_i$)$^{0(+)}$, the N--H distance is 1.00 {\AA (1.02 {\AA}) and the distance between Eu and the N atom in the NH unit is 2.34 {\AA} (2.30 {\AA}); see Fig.~\ref{fig;struct3}(a). 

Eu$_{\rm Ga}$-C$_{\rm N}$ has two geometric configurations associated with two possible positions of C$_{\rm N}$ with respect to Eu$_{\rm Ga}$. Eu$_{\rm Ga}$-C$_{\rm N}$ introduces two defect levels: $(+/0)$ at 0.33 eV (0.17 eV) above the VBM and $(0/-)$ at 1.08 eV (1.17 eV) below the CBM for the basal (axial) configuration. The distance between the two defects in the complex is 2.27 {\AA}, 2.35 {\AA}, or 2.23 {\AA} for the $0$, $+$, or $-$ charge state, respectively; see Fig.~\ref{fig;struct3}(b). (Eu$_{\rm Ga}$-C$_{\rm N}$)$^-$ has a negative binding energy and is thus unstable toward its isolated constituents, Eu$_{\rm Ga}^0$ and C$_{\rm N}^-$. In other words, Eu$_{\rm Ga}$-C$_{\rm N}$ is unlikely to be stable as a defect complex when incorporated under n-type conditions. The other stable charge states of the defect complex have positive but small binding energies; see Table \ref{tab;complex}. A combination of such low calculated binding energies and high formation energies (see Fig.~\ref{fig;fe3}) suggests that Eu$_{\rm Ga}$-C$_{\rm N}$ is unlikely to be stable as a complex under thermodynamic equilibrium. 

Eu$_{\rm Ga}$-O$_{\rm N}$ has one defect level in the bulk band gap: $(+/0)$ is at 1.26 eV (1.23 eV) below the CBM for the basal (axial) geometric configuration. The basal configuration is lower in energy than the axial one (not included in Fig.~\ref{fig;fe3}) by 0.18 eV (0.15 eV) when in its $0$ ($+$) charge state. (Eu$_{\rm Ga}$-O$_{\rm N}$)$^0$ is a complex of Eu$_{\rm Ga}^-$ and O$_{\rm N}^+$, whereas (Eu$_{\rm Ga}$-O$_{\rm N}$)$^+$ is a complex of Eu$_{\rm Ga}^0$ and O$_{\rm N}^+$. In the basal configuration, the Eu--O distance is 2.29 {\AA} (2.23 {\AA}) in the $0$ ($+$) charge state; see Fig.~\ref{fig;struct3}(c). Eu$_{\rm Ga}$-O$_{\rm N}$ can have a much lower formation energy than the unassociated Eu$_{\rm Ga}$, as seen in Fig.~\ref{fig;fe3}. Similarly, Eu$_{\rm Ga}$-Si$_{\rm Ga}$ introduces the $(+/0)$ level at 1.04 eV below the CBM. (Eu$_{\rm Ga}$-Si$_{\rm Ga}$)$^0$ is a complex of Eu$_{\rm Ga}^-$ and Si$_{\rm Ga}^+$, whereas (Eu$_{\rm Ga}$-Si$_{\rm N}$)$^+$ is a complex of Eu$_{\rm Ga}^0$ and Si$_{\rm Ga}^+$. The Eu--Si distance is 3.25 {\AA} (3.27 {\AA}) in the $0$ ($+$) charge state; see Fig.~\ref{fig;struct3}(d). 

The electronic behavior of Eu$_{\rm Ga}$-C$_{\rm N}$ is thus similar to that of Er$_{\rm Ga}$-C$_{\rm N}$ \cite{Hoang2015RRL}. Other defect complexes are different. For example, Er$_{\rm Ga}$-O$_{\rm N}$ is a shallow donor and thus has no defect levels in the bulk band gap; Er$_{\rm Ga}$-H$_i$ introduces the $(+/-)$ level \cite{Hoang2015RRL} instead of $(+/0)$ like in the case of Eu$_{\rm Ga}$-H$_i$. This, again, illustrates the difference between Er and the mixed-valence Eu in GaN. 

\subsection{Defect complexes of Eu, O, and Mg}\label{sec;withmg}

\begin{figure}[t]%
\vspace{0.2cm}
\includegraphics*[width=\linewidth]{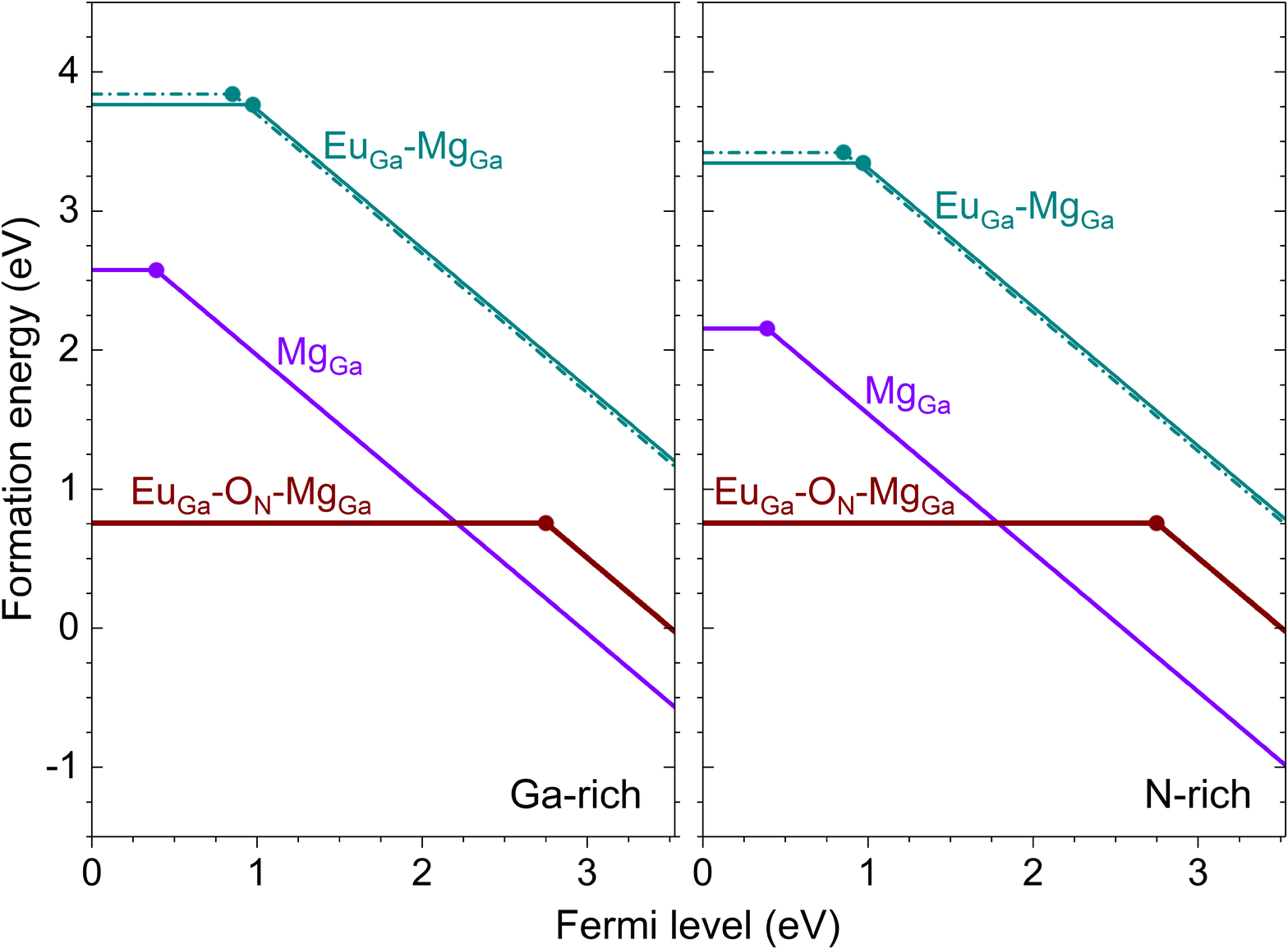}
\caption{Formation energies of Mg$_{\rm Ga}$ and related defect complexes Eu$_{\rm Ga}$-Mg$_{\rm Ga}$ and Eu$_{\rm Ga}$-O$_{\rm N}$-Mg$_{\rm Ga}$ in GaN, plotted as a function of Fermi level from the VBM to the CBM, under the Ga-rich and N-rich conditions. For each defect, only segments corresponding to the lowest-energy charge states are shown. Large solid dots connecting two segments with different slopes mark the defect levels. Two Eu$_{\rm Ga}$-Mg$_{\rm Ga}$ configurations, corresponding the structures in Fig.~\ref{fig;struct4}(a) and \ref{fig;struct4}(b), are reported.}
\label{fig;fe4}
\end{figure}

\begin{figure}[t]%
\vspace{0.2cm}
\includegraphics*[width=\linewidth]{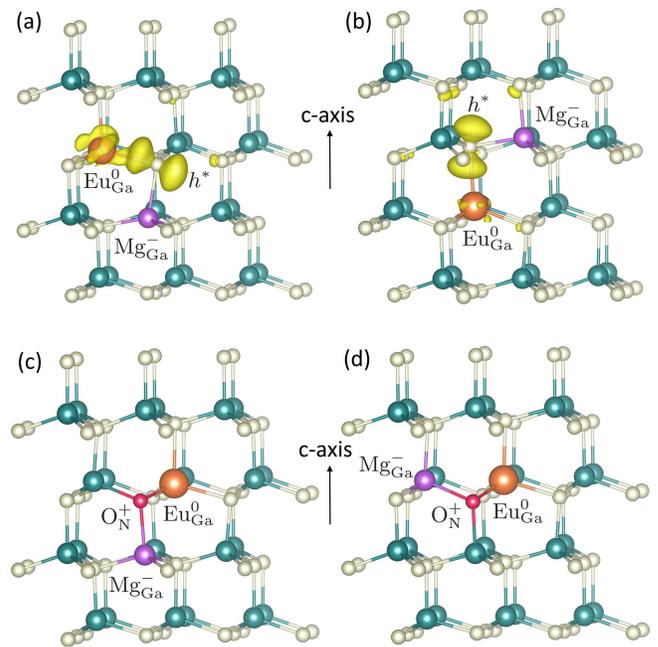}
\caption{Structure of Mg-related defect complexes: (Eu$_{\rm Ga}$-Mg$_{\rm Ga}$)$^0$ [(a) and (b)] and (Eu$_{\rm Ga}$-O$_{\rm N}$-Mg$_{\rm Ga}$)$^0$ [(c) and (d)]. Large spheres are Eu, medium Ga/Mg, and small N/O. Charge densities associated with the localized hole $h^\ast$ in the (Eu$_{\rm Ga}$-Mg$_{\rm Ga}$)$^0$ configuration are visualized as (yellow) isosurfaces; the isovalue for the isosurface is set to 0.05 $e$/{\AA}$^{3}$.}
\label{fig;struct4}
\end{figure}

We now focus on possible interaction between Eu, Mg, and O in GaN. Figure \ref{fig;fe4} shows the calculated formation energy of Mg-related defects. The unassociated Mg$_{\rm Ga}$ has the $(0/-)$ level at 0.39 eV above the VBM, in reasonable agreement with previous studies~\cite{Lany2010APL,Lyons2012}. It is noted that ``Mg$_{\rm Ga}^0$'' is not a true charge state of Mg$_{\rm Ga}$, but a defect complex consisting of Mg$_{\rm Ga}^-$ and an electron hole ($h^\ast$) localized on one of the basal N atoms. The Mg--N distance is 2.22 {\AA} for the N atom that hosts $h^\ast$ and 2.00--2.01 {\AA} for the other N atoms. The axial configuration of Mg$_{\rm Ga}^0$ is 10 meV higher in energy than the basal one. A metastable configuration of Mg$_{\rm Ga}^0$ in which the hole is delocalized over {\it all} N atoms is 0.19 eV higher in energy than the ground-state one. In this configuration, all the Mg--N distances are almost equal (2.02--2.04 {\AA}).  %energy barrier?   

Eu$_{\rm Ga}$-Mg$_{\rm Ga}$ has a defect level, $(0/-)$, at 0.97 eV or 0.85 eV above the VBM, see Fig.~\ref{fig;fe4}, depending on specific geometric configurations. (Eu$_{\rm Ga}$-Mg$_{\rm Ga}$)$^0$ is a complex consisting of Eu$_{\rm Ga}^0$, Mg$_{\rm Ga}^-$, and $h^\ast$, whereas (Eu$_{\rm Ga}$-Mg$_{\rm Ga}$)$^-$ is a complex of Eu$_{\rm Ga}^0$ and Mg$_{\rm Ga}^-$. Figure~\ref{fig;struct4}(a) shows the lowest-energy configuration of (Eu$_{\rm Ga}$-Mg$_{\rm Ga}$)$^0$ in which $h^\ast$ resides on the N atom that is basally (axially) bonded to Eu (Mg). Other configurations, such as that shown in Fig.~\ref{fig;struct4}(b), are 38--75 meV higher in energy. In the basal (with respect to Eu) configuration, see Figure~\ref{fig;struct4}(a), the distance between Mg and the N site that bridges Mg and Eu is 2.12 {\AA} (2.02 {\AA}) when the complex is in the $0$ ($-$) charge state. The presence of $h^\ast$ on that N atom thus slightly elongates the Mg--N bond. The Eu--N distance is also longer for the N atom that hosts $h^\ast$ [2.23 {\AA}, compared to 2.25 {\AA} (2.18 {\AA}) for the axial (other basal) Eu--N bonds in (Eu$_{\rm Ga}$-Mg$_{\rm Ga}$)$^0$]. In the axial configuration, see Figure~\ref{fig;struct4}(b), the distance between Mg and the N site that bridges Mg and Eu is also longer when $h^\ast$ is on that N atom: the Mg--N bond length is 2.09 {\AA} (2.01 {\AA}) in (Eu$_{\rm Ga}$-Mg$_{\rm Ga}$)$^{0(-)}$. The calculated binding energy of (Eu$_{\rm Ga}$-Mg$_{\rm Ga}$)$^-$ is almost zero, suggesting that Eu$_{\rm Ga}$-Mg$_{\rm Ga}$ may be not stable as a complex when incorporated under n-type, thermodynamic equilibrium growth conditions (see also discussion in Sec.~\ref{sec;dis1}). 

Note that, unlike Mg$_{\rm Ga}^0$, a metastable state of (Eu$_{\rm Ga}$-Mg$_{\rm Ga}$)$^0$ in which the hole is delocalized over the N atoms cannot be stabilized (even at the DFT-GGA \cite{GGA} level of the calculations where the electronic states tend to be over-delocalized). This is due to the local lattice distortion caused by the presence of Eu in the complex.   

Finally, we consider a complex consisting of Eu$_{\rm Ga}$, O$_{\rm N}$, and Mg$_{\rm Ga}$. Eu$_{\rm Ga}$-O$_{\rm N}$-Mg$_{\rm Ga}$ introduces one defect level in the band gap: $(0/-)$ at 0.78 eV below the CBM; see Fig.~\ref{fig;fe4}. The neutral charge state, (Eu$_{\rm Ga}$-O$_{\rm N}$-Mg$_{\rm Ga}$)$^0$, is a defect complex consisting of Eu$_{\rm Ga}^0$, O$_{\rm N}^+$, and Mg$_{\rm Ga}^-$, whereas (Eu$_{\rm Ga}$-O$_{\rm N}$-Mg$_{\rm Ga}$)$^-$ is a complex of Eu$_{\rm Ga}^-$, O$_{\rm N}^+$, and Mg$_{\rm Ga}^-$; see also Table \ref{tab;complex}. Figures~\ref{fig;struct4}(c) and \ref{fig;struct4}(d) show the two lowest configurations of the neutral charge state which have almost equal energies with the former 10 meV lower in energy than the latter. Both charge states can have significantly lower formation energies than those of the other defects and relatively high binding energies.

\section{Discussion}\label{sec;dis}

\subsection{Tuning the Eu valence and concentration through co-doping and defect association}\label{sec;dis1}

\begin{figure}[t]%
\vspace{0.2cm}
\includegraphics*[width=0.99\linewidth]{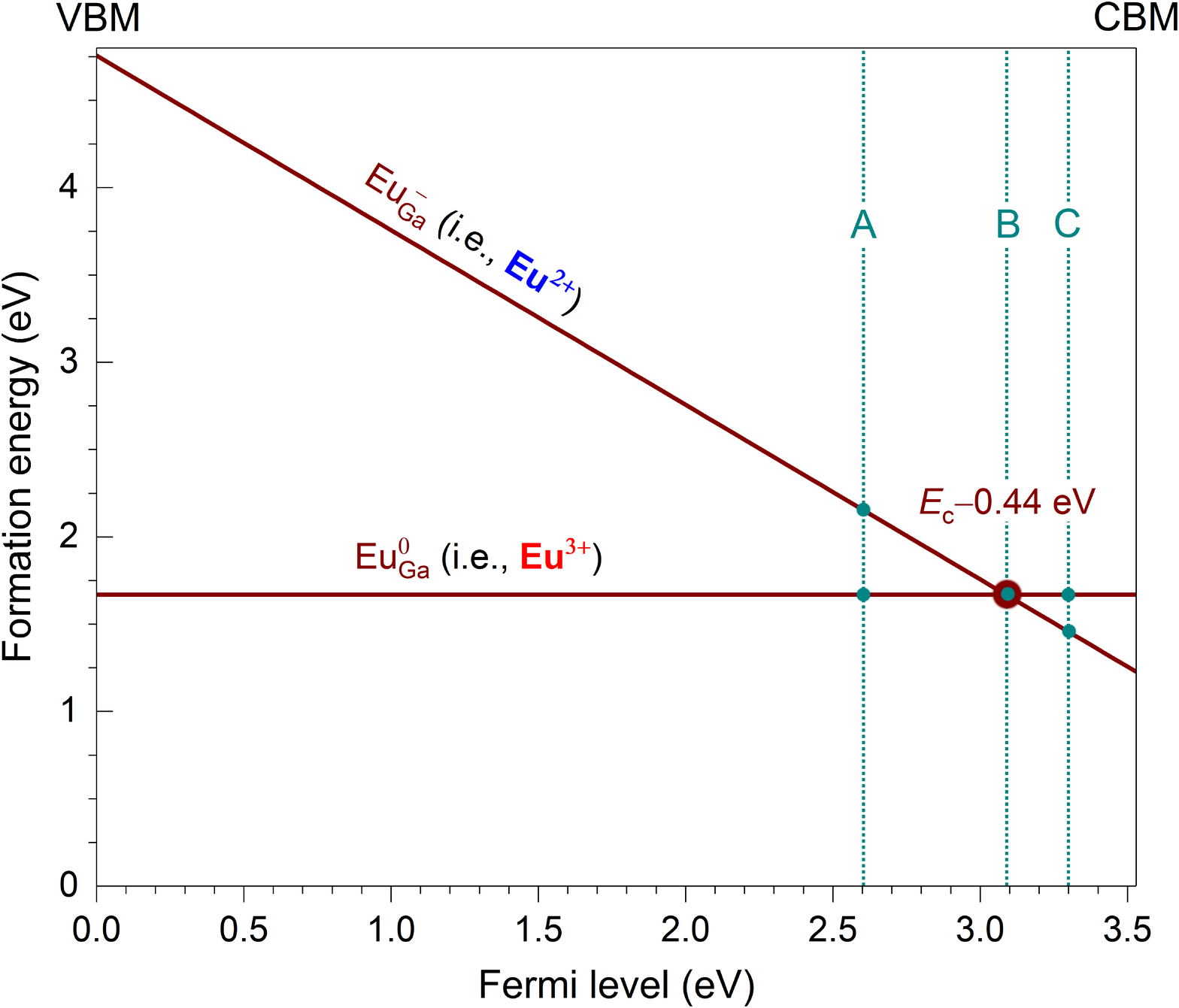}
\caption{The formation energy of the negatively charged configuration of Eu$_{\rm Ga}$ is dependent on the Fermi-level position. This allows for the tuning of the Eu$^{2+}$/Eu$^{3+}$ ratio by shifting the Fermi level, e.g., through co-doping; see the text.}
\label{fig;fermi}
\end{figure}

Our results clearly show that Eu can be stable as Eu$^{2+}$ and/or Eu$^{3+}$ in GaN. Figure \ref{fig;fermi} highlights the dependence of the Eu$^{2+}$/Eu$^{3+}$ ratio on the position of Fermi level ($\mu_e$). For $\mu_e$ $<$ $E_c-0.44$ eV, the formation energy of Eu$_{\rm Ga}^0$ is lower than that of Eu$_{\rm Ga}^-$. At position A, for example, the concentration of Eu$_{\rm Ga}^0$ is much larger than that of Eu$_{\rm Ga}^-$ and hence $c({\rm Eu}^{2+})/c({\rm Eu}^{3+})\ll 1$. At $\mu_e$ $=$ $E_c-0.44$ eV (i.e., position B), Eu$_{\rm Ga}^0$ and Eu$_{\rm Ga}^-$ have equal concentrations; i.e., $c({\rm Eu}^{2+})/c({\rm Eu}^{3+}) = 1$. For $\mu_e$ $>$ $E_c-0.44$ eV, Eu$_{\rm Ga}^-$ is energetically more stable than Eu$_{\rm Ga}^0$; at position C, e.g., $c({\rm Eu}^{2+})/c({\rm Eu}^{3+}) \gg 1$. Even a small shift in the Fermi-level position (hence a small change in the formation energy of Eu$_{\rm Ga}^-$) can lead to a large change in the Eu$^{2+}$ concentration; see Eq.~(\ref{eq;con}).   

As shallow donors, O and Si can make GaN n-type or, at least, shift the Fermi level toward the CBM as the charge neutrality condition [Eq.~(\ref{eq:neutrality})] is re-established. When the Fermi level moves closer to the CBM, the n-type carrier concentration increases. In the case of Eu-doped GaN, O and/or Si co-doping can be employed to control the charge state of Eu$_{\rm Ga}$ and thus the valence state of Eu. For example, with an appropriate concentration of the co-dopants, the Fermi level of the material can be ``pinned'' near or above $E_c-0.44$ eV where the concentration of Eu$^{2+}$ is high (See below for a discussion of relevant reported experiments). Donor-like defects such as the positively charged configurations of $V_{\rm N}$ and H$_i$ can have similar effects, although they are expected to be much less effective than the shallow donors in shifting the Fermi level. We emphasize that these are {\it global} effects since those defects do not need to be close to Eu$_{\rm Ga}$ for the Fermi-level shift to happen; the defect--defect interaction takes place only indirectly via the interaction with the common electron reservoir [i.e., $\mu_e$ in Eq.~(\ref{eq:eform})]. 

The effects of defect association and thus {\it local} defect--defect interaction are investigated by considering defect complexes explicitly, as presented in Sec.~\ref{sec;results}. Overall, we find that the electronic behavior of complexes is significantly different from that of their unassociated constituents. Defect levels associated with the complexes are shifted, compared to those associated with the isolated ones, and additional levels may form as a result of strong local elastic and electrostatic interactions between constituents in the complexes. Notably, the valence state of Eu can be controlled through defect association. For example, Eu$^{2+}$ is found to be more stable in complexes with $V_{\rm N}$, H$_i$, O$_{\rm N}$, and Si$_{\rm Ga}$, compared to that in the unassociated Eu$_{\rm Ga}$. This is mainly due to the strong Coulomb attraction between Eu$_{\rm Ga}^-$ and the positively charged native defect in the neutral complex configuration (See a detailed discussion in the case of Eu$_{\rm Ga}$-$V_{\rm N}$ in Sec.~\ref{sec;withnative}). In complexes with C$_{\rm N}$ and Mg$_{\rm Ga}$, Eu is stable only as Eu$^{3+}$, largely determined by the available stable charge states of C$_{\rm N}$ and Mg$_{\rm Ga}$. Defect association also changes the formation energy and can thus affect defect incorporation during growth. Complexes of Eu$_{\rm Ga}$ and O$_{\rm N}$ can have a much lower formation energy than the unassociated Eu$_{\rm Ga}$, indicating that co-doping with O makes it easier to incorporate Eu into GaN. The formation energy is significantly lower in the case of Eu$_{\rm Ga}$-O$_{\rm N}$-Mg$_{\rm Ga}$. 

The results summarized in Table \ref{tab;complex} also show that the binding energy varies significantly from one defect complex to another. Having a positive calculated binding energy, however, does not mean that the defect complex will readily form. As discussed in Ref.~\cite{walle:3851}, under thermodynamic equilibrium, the binding energy needs to be greater than the larger of the formation energies of the isolated constituent defects for the complex to have higher concentration than its constituents. On the other hand, a small calculated binding energy does not necessarily mean that the complex cannot occur with a significant concentration since it can still form under non-equilibrium growth conditions and get trapped inside the material. (Eu,Mg)-doped GaN, for example, has often been made by ion-implanting with Eu fluences \cite{ODonnell2016APL,Singh2017SC} or prepared by molecular-beam expitaxy (MBE) \cite{Sekiguchi2013JAP,Sekiguchi2016APL}. Defect complexes such as Eu$_{\rm Ga}$-Mg$_{\rm Ga}$ thus can still exist despite having a very small binding energy. Overall, one should expect that defects are present in the material in both the unassociated and associated forms. 

Our results for the unassociated Eu$_{\rm Ga}$ showing Eu$^{2+}$ more stable than Eu$^{3+}$ only in a small range of Fermi-level values near the CBM, see Fig.~\ref{fig;fe1} or \ref{fig;fermi}, thus explain why Eu$^{3+}$ is often found to be predominant in Eu-doped GaN samples. Experimentally, a significant Eu$^{2+}$ concentration occurs only when prepared under certain conditions \cite{Mitchell2017MCP,Nunokawa2020JAP}. Mitchell et al.~\cite{Mitchell2017MCP} were able to achieve $c({\rm Eu}^{2+})/c({\rm Eu}^{3+}) > 1$ when using both O and Si as co-dopants and suitable growth conditions. Nunokawa et al.~\cite{Nunokawa2020JAP} reported similar achievement with their (Eu,Si,O)-doped GaN. Notably, all their samples exhibited n-type conductivity \cite{Nunokawa2020JAP}, which is consistent with our results showing the $\epsilon(0/-)$ level of Eu$_{\rm Ga}$, i.e., the Eu$^{3+/2+}$ level, close to the CBM. A measurement of the Fermi level as a function of the O and Si concentration should be able to confirm our prediction of this defect level. Note that the Eu-doped GaN/SiO$_2$ nanocomposites reported in Ref.~\cite{Mahalingam2007AFM} can be regarded as Eu-doped GaN being co-doped with O and Si, and Eu$^{2+}$ is expected to be present predominantly at/near the interface. 

The relatively low temperature (e.g., 700$^{\circ}$C instead of 1030$^{\circ}$C) used during growth via OMVPE also appeared to play a key role in increasing the Eu$^{2+}$ concentration \cite{Zhu2016APLM,Mitchell2017MCP,Nunokawa2020JAP}. This is likely because decreasing the temperature leads to an increase in the concentration of complexes and a decrease in the concentration of their isolated constituents; see a discussion in Ref.~\cite{walle:3851}. In other words, it is easier to incorporate Eu into GaN in the form of complexes such as the low-formation-energy Eu$_{\rm Ga}$-O$_{\rm N}$ at lower growth temperatures. The temperature, of course, cannot be too low as the concentration of thermally activated defects is still governed by Eq.~(\ref{eq;con}). 

\subsection{Eu-related defects as carrier traps for Eu$^{3+}$ intra-$f$ luminescence}\label{sec;dis2}

We now focus on the electronic behavior of the Eu-related defects and discuss their possible role in intra-$f$ luminescence through non-resonant excitation of the Eu$^{3+}$ ion; see Fig.~\ref{fig;excitation}. As reported earlier, the unassociated Eu$_{\rm Ga}$ has defect levels in the host band gap. The $(0/-)$ level can act as an electron trap. The electron-capturing defect configuration is Eu$_{\rm Ga}^0$, which is essentially Eu$^{3+}$. When Eu$_{\rm Ga}^0$ captures an electron from the conduction band (e.g., previously excited from the valence band to the conduction band under band-to-band excitation) and becomes Eu$_{\rm Ga}^-$ (assuming that the system has enough time to relax to its equilibrium configuration), the valence state of Eu changes from trivalent to divalent. The captured electron in Eu$_{\rm Ga}^-$ then recombines non-radiatively with a free hole from the valence band or a hole at some acceptor level and transfer the recombination energy into the Eu$^{3+}$ $4f$-core. The $(+/0)$ level can act as a hole trap. Compared to Eu-related complexes discussed below, the unassociated Eu$_{\rm Ga}$ traps are expected to be less effective since there is no Coulomb attraction between the carrier and the carrier-capturing configuration (Eu$_{\rm Ga}^0$). Note that the carrier capture cross section can decrease by orders of magnitude in going from Coulomb attractive defect centers (e.g., $\sim$$10^{-12}$--$10^{-15}$ cm$^{-2}$) to neutral centers ($\sim$$10^{-15}$--$10^{-17}$ cm$^{-2}$) to repulsive centers ($\sim$$10^{-22}$ cm$^{-2}$) \cite{Mayer1990}. The relatively small energy separation between the defect level and the band edge may also increase the likelihood of the captured carrier being thermally re-excited into the band.

Experimentally, the unassociated Eu$_{\rm Ga}$ is believed by many to be the dominant Eu$^{3+}$ center in Eu-doped GaN samples \cite{Roqan2010PRB,ODonnell2011OM,Woodward2011OM}. The luminescence center is often characterized by its high relative abundance (up to more than 97\% of the incorporated Eu) \cite{Woodward2011APL,Wakamatsu2013JAP}, low-efficiency energy transfer from the GaN host into the Eu$^{3+}$ $4f$-core (the effective excitation cross section $\sim$$1.2\times 10^{-17}$ cm$^2$) \cite{Woodward2011APL,Wakamatsu2013JAP,Timmerman2020PRA}, and strong thermal quenching \cite{Timmerman2020PRA}. These descriptions appear to be consistent with the characteristics of the unassociated Eu$_{\rm Ga}$ center reported in this work.      

For Eu$_{\rm Ga}$-$V_{\rm N}$, the $(+/0)$ level can act as a deep electron trap. An electron is likely to be captured at the $V_{\rm N}^+$ part of the carrier-capturing defect configuration (Eu$_{\rm Ga}$-$V_{\rm N}$)$^+$, due to the Coulomb attraction. This trap is 1.86 eV above the VBM, slightly smaller than the separation (2.12 eV) \cite{Fleischman2009APB} between the $^7F_0$ and $^5D_0$ levels of Eu$^{3+}$. Note, however, that the error bar in our calculation of the defect level is about 0.1 eV; thus the trap may actually be slightly higher and the energy obtained from a non-radiative recombination of the trapped electron at the $(+/0)$ level and a hole may be large enough to excite an electron from $^7F_0$ to $^5D_0$. In other words, we cannot completely rule out the role of Eu$_{\rm Ga}$-$V_{\rm N}$ as an electron trap and a defect center for the $^5D_J$ $\rightarrow$ $^7F_J$ transitions, although it is very likely that the complex has a limited role in high-energy luminescent transitions. The defect levels nearer to the VBM, $(3+/2+)$ and $(2+/+)$, may act as hole traps; however, their hole-capture efficiency should be very low given the Coulomb repulsion between the carrier and the positively charged ($V_{\rm N}^{3+}$ or $V_{\rm N}^+$) part in the hole-capturing defect configuration of Eu$_{\rm Ga}$-$V_{\rm N}$. 

For Eu$_{\rm Ga}$-$V_{\rm Ga}$, the defect levels nearer to the VBM and CBM may act as hole and electron traps, respectively. However, they are unlikely to be efficient due to the Coulomb repulsion between the carrier and certain parts of the carrier-capturing defect configurations; see Fig.~\ref{fig;struct2}. The traps formed by Eu$_{\rm Ga}$-C$_{\rm N}$ are also expected not to be very efficient because the carrier-capturing defect configuration (Eu$_{\rm Ga}$-C$_{\rm N}$)$^0$ is all neutral; see Fig.~\ref{fig;struct3}(b). 

The $(+/0)$ level of Eu$_{\rm Ga}$-H$_i$, Eu$_{\rm Ga}$-O$_{\rm N}$, and Eu$_{\rm Ga}$-Si$_{\rm Ga}$ is expected to be an efficient deep electron trap. An excited electron from the conduction band is likely to be captured at the positively charged (H$_i^+$, O$_{\rm N}^+$, or Si$_{\rm Ga}^+$) part of the carrier-capturing defect configuration, due to the Coulomb attraction; see Fig.~\ref{fig;struct3}(a), \ref{fig;struct3}(c), and \ref{fig;struct3}(d). In the case of Eu$_{\rm Ga}$-O$_{\rm N}$, for example, since the neutral charge state of O$_{\rm N}$ is energetically unstable, the captured electron is likely transferred to the Eu$_{\rm Ga}^0$ part and transforms it into Eu$_{\rm Ga}^-$, assuming the system is allowed to relax to its equilibrium configuration, and hence (Eu$_{\rm Ga}$-O$_{\rm N}$)$^+$ becomes (Eu$_{\rm Ga}$-O$_{\rm N}$)$^0$ which then recombines with a hole and the energy is transferred into the Eu $4f$-core.      

Regarding the Mg-containing defect complexes, the $(0/-)$ level of Eu$_{\rm Ga}$-Mg$_{\rm Ga}$ can act as a deep hole trap. A hole from the valence band (e.g., previously created by exciting an electron from the valence band to the conduction band) can be efficiently captured at the negatively charged (Mg$_{\rm Ga}^-$) part of the carrier-capturing defect configuration (Eu$_{\rm Ga}$-Mg$_{\rm Ga}$)$^-$; see Table \ref{tab;complex}. Finally, the $(0/-)$ level of Eu$_{\rm Ga}$-O$_{\rm N}$-Mg$_{\rm Ga}$ can act as an electron trap with the carrier likely being captured at the O$_{\rm N}^+$ part of (Eu$_{\rm Ga}$-O$_{\rm N}$-Mg$_{\rm Ga}$)$^0$. The behavior of this neutral configuration should be similar to that of (Eu$_{\rm Ga}$-O$_{\rm N}$)$^+$ described above.  

Altogether we find that Eu$_{\rm Ga}$-O$_{\rm N}$, Eu$_{\rm Ga}$-Si$_{\rm Ga}$, Eu$_{\rm Ga}$-H$_i$, Eu$_{\rm Ga}$-Mg$_{\rm Ga}$, Eu$_{\rm Ga}$-O$_{\rm N}$-Mg$_{\rm Ga}$, and possibly Eu$_{\rm Ga}$-$V_{\rm N}$ are efficient defect-related Eu$^{3+}$ centers for non-resonant excitation. The significant local distortion around the Eu$^{3+}$ ion should help relax the Laporte selection rules and allows for bright emission. As discussed above, they are efficient carrier traps and thus likely to have high carrier capture cross sections. The energy transfer from the GaN host into the Eu$^{3+}$ $4f$-core is also expected to be efficient, given the close proximity of the carrier-capturing part to the Eu$^{3+}$ ion in the carrier-capturing defect configuration. Experimentally, it was shown that efficient energy transfer into the Eu $4f$-core and a high concentration of the Eu-related defect centers is key to enhanced emission intensity \cite{Inaba2018JAP}. Our findings are thus consistent with experimental observations showing that the Eu$^{3+}$ PL emission was significantly enhanced in GaN co-doped with Eu and O \cite{Mitchell2016SR,Zhu2016APLM,Mitchell2017MCP}, Si \cite{Wang2009JAP}, or Mg \cite{Takagi2011APL,Lee2012APL,Wakahara2012JL,Sekiguchi2013JAP,Sekiguchi2016APL}.

We now comment on the PL hysteresis observed in (Eu,Mg)-doped GaN, believed to involve hysteretic photochromic switching (HPS) between two defect configurations, namely ``Eu0'' and ``Eu1(Mg)'', in which the Eu$^{3+}$ ion experiences slightly different local crystal fields \cite{ODonnell2014PSSC,ODonnell2016APL,Singh2017SC}. The authors identified ``Eu0'' and ``Eu1(Mg)'' with the so-called ``shallow transient state'' (STS) and ``deep ground state'' (DGS), respectively, proposed by Lany and Zunger~\cite{Lany2010APL} for Mg$_{\rm Ga}^0$ in Mg-doped GaN. The DGS is equivalent to the Mg$_{\rm Ga}^0$ configuration consisting of Mg$_{\rm Ga}^-$ and $h^\ast$ in our work and the STS can be identified with the metastable configuration described in Sec.~\ref{sec;withmg}. One assumption made by O'Donnell et al.~\cite{ODonnell2016APL} was that, in (Eu,Mg)-doped GaN, Eu could be regarded as a ``spectator ion''. This may not be the case as we find that, e.g., the $(0/-)$ level of Eu$_{\rm Ga}$-Mg$_{\rm Ga}$ is shifted by $\sim$0.5 eV from that of Mg$_{\rm Ga}$ and the local lattice environment and hence the ability to accommodate a metastable state is different for (Eu$_{\rm Ga}$-Mg$_{\rm Ga}$)$^0$ and Mg$_{\rm Ga}^0$; see Sec.~\ref{sec;withmg}. Note that Mg in (Eu,Mg)-doped GaN is expected to be present both as the unassociated Mg$_{\rm Ga}$ and in Eu$_{\rm Ga}$-Mg$_{\rm Ga}$. And it is not clear at this point if the perturbation is strong enough to cause the observed PL hysteresis when Mg$_{\rm Ga}$ and Eu$_{\rm Ga}$ are far apart such that the metastable configuration of Mg$_{\rm Ga}^0$ can be stabilized. Besides, we find the total-energy difference between the stable and metastable configurations of Mg$_{\rm Ga}^0$ is rather large (0.19 eV). 

In (Eu,O,Mg)-doped GaN, Cameron et al.~\cite{Cameron2020APL} observed another Eu$^{3+}$ center denoted as ``Eu0(Ox)'' in addition to ``Eu1(Mg)'' and ``Eu0''. Eu0(Ox) was found to be stable over a prolonged excitation time and a wide temperature range, unlike the other two Eu$^{3+}$ centers. The center can be identified with the Eu$_{\rm Ga}$-O$_{\rm N}$-Mg$_{\rm Ga}$ complex in our calculations which should co-exist with smaller defect complexes such as Eu$_{\rm Ga}$-Mg$_{\rm Ga}$ and the unassociated defects. With its low formation energy, the complex is expected to occur with a significant concentration. The main difference between Eu$_{\rm Ga}$-Mg$_{\rm Ga}$ and Eu$_{\rm Ga}$-O$_{\rm N}$-Mg$_{\rm Ga}$ is that upon capturing a hole the negatively charged state of the former becomes (Eu$_{\rm Ga}$-Mg$_{\rm Ga}$)$^0$ with a localized hole residing at the bridging N atom, whereas upon capturing an electron the neutral state of the latter becomes (Eu$_{\rm Ga}$-O$_{\rm N}$-Mg$_{\rm Ga}$)$^-$ with the valence change occurring on the Eu ion; all assuming the system is allowed to relax to its equilibrium configuration. 

\subsection{Eu-related defects as carrier traps for defect-to-band luminescence}\label{sec;dis3} 

\begin{figure*}[t]%
\vspace{0.2cm}
\includegraphics[width=17.5cm]{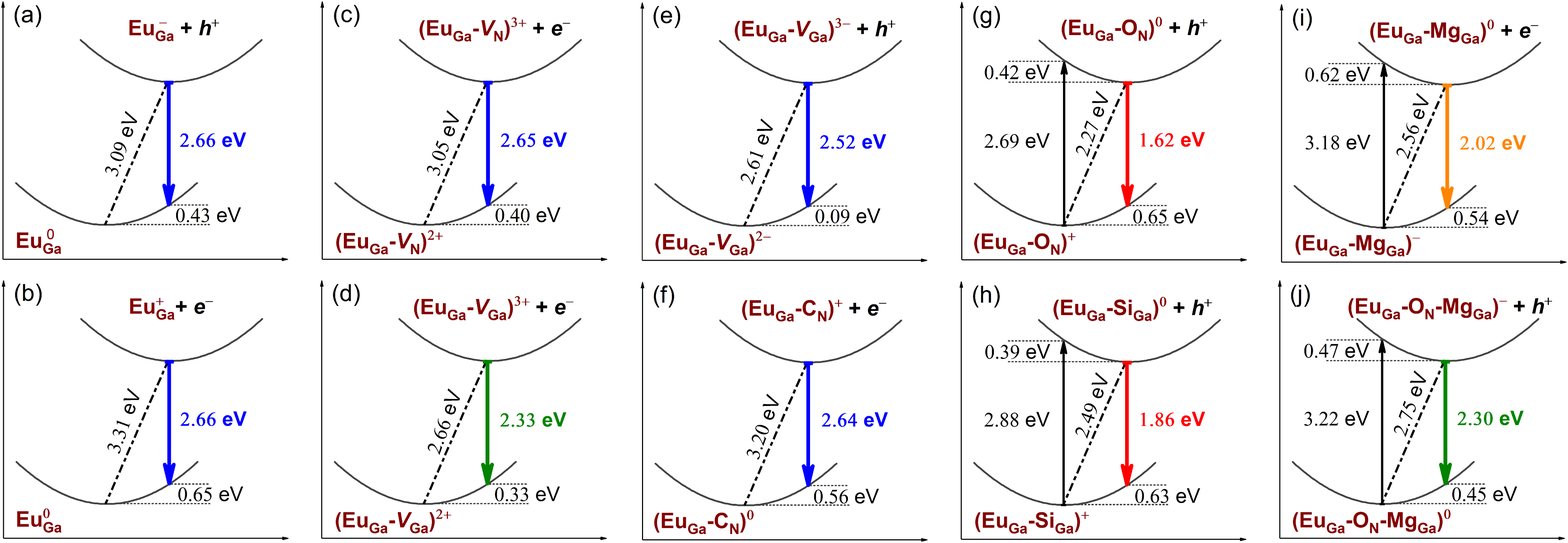}
\caption{Configuration-coordinate diagrams illustrating optical emission (down arrow) and absorption (up arrow) processes involving Eu-related defects in GaN. The thermal energy [also called the zero-phonon line (ZPL); the dash-dotted line] is the thermodynamic transition level $\epsilon(q/q')$ relative to the VBM [in the case of hole ($h^+$) capture) or CBM [electron ($e^-$) capture]. The values sandwiched between two dotted lines are the relaxation energies (the Franck-Condon shifts). Only transitions with emissions in the visible range and the absorption peaks that fall within the host band gap are included. Axes are not to scale.} 
\label{fig;cc}
\end{figure*}

In addition to intra-$f$ luminescence, Eu-related defects can also act as carrier traps in defect-to-band transitions that emit light in the visible range. In n-type GaN, the unassociated Eu$_{\rm Ga}$ is stable as Eu$_{\rm Ga}^-$. This defect configuration can capture a hole from the VBM (e.g., left by an electron previously excited to from the valence to the conduction band) and emits a photon. The peak emission energy corresponds to the transition level $E_{\rm opt}^{-/0}$, i.e., the energy difference between Eu$_{\rm Ga}^-$ and Eu$_{\rm Ga}^0$ in the lattice configuration of Eu$_{\rm Ga}^-$. As shown in Fig.~\ref{fig;cc}(a), this emission peak is at 2.66 eV, with a relaxation energy of 0.43 eV. Our results thus indicate that Eu$_{\rm Ga}$ is a source of blue luminescence in n-type GaN, which may explain the broad blue emission observed in the Eu$^{2+}$-containing GaN/SiO$_2$ nanocomposites \cite{Mahalingam2007AFM}. It should be emphasized that the mechanism we present here is different from that associated with the so-called $4f^65d^1 \rightarrow 4f^7$ transition within the Eu$^{2+}$ ion mentioned in Ref.~\cite{Mahalingam2007AFM} and often regarded in the literature as the cause for blue luminescence observed in Eu$^{2+}$-containing materials. In p-type GaN, Eu$_{\rm Ga}$ is stable as Eu$_{\rm Ga}^+$; the emission peak corresponds to the Eu$_{\rm Ga}^+$ $\rightarrow$ Eu$_{\rm Ga}^0$ transition, after capturing an electron from the CBM, is also 2.66 eV (blue), see Fig.~\ref{fig;cc}(b).         

Similar calculations are carried out for Eu-related defect complexes, as shown in Figs.~\ref{fig;cc}(c)--\ref{fig;cc}(j) . In n-type GaN, for example, (Eu$_{\rm Ga}$-$V_{\rm Ga}$)$^{3-}$ can give rise to broad emission peaked at 2.52 eV (blue), (Eu$_{\rm Ga}$-O$_{\rm N}$)$^0$ at 1.62 eV (red), (Eu$_{\rm Ga}$-Si$_{\rm Ga}$)$^0$ at 1.86 eV (red), and (Eu$_{\rm Ga}$-O$_{\rm N}$-Mg$_{\rm Ga}$)$^-$ at 2.30 eV (green), after capturing a hole from the VBM. Each of these defect configurations is the most stable charge state of its respective defect under n-type conditions; see Sec.~\ref{sec;results}. In p-type GaN, (Eu$_{\rm Ga}$-$V_{\rm N}$)$^{3+}$ can give rise to broad emission peaked at 2.65 eV (blue), (Eu$_{\rm Ga}$-$V_{\rm Ga}$)$^{3+}$ at 2.33 eV (green), (Eu$_{\rm Ga}$-C$_{\rm N}$)$^+$ at 2.64 eV (blue), and (Eu$_{\rm Ga}$-Mg$_{\rm Ga}$)$^0$ at 2.02 eV (orange), after capturing an electron from the CBM. Each of these defect configurations is the most stable charge state of its respective defect under p-type conditions. We also consider the absorption process. In the case of Eu$_{\rm Ga}$-Mg$_{\rm Ga}$, for example, an electron can be excited from (Eu$_{\rm Ga}$-Mg$_{\rm Ga}$)$^-$ to the conduction band, with a peak absorption energy of 3.18 eV, given by the energy difference between (Eu$_{\rm Ga}$-Mg$_{\rm Ga}$)$^-$ and (Eu$_{\rm Ga}$-Mg$_{\rm Ga}$)$^0$, both in the lattice configuration of the negatively charged configuration.  

The above-mentioned Eu-related defects can thus be sources of broad blue, green, red, and orange defect-to-band luminescence in n- or p-type GaN samples.

\section{Conclusions} 

We have carried out a systematic study of defects in Eu-doped GaN using hybrid density-functional calculations. The material is found to exhibit rich defect physics resulting from the ability of Eu to be mixed-valence and the interaction between the RE dopant and native defects and (intentional or otherwise) impurities. Eu can be stable as divalent and/or trivalent when incorporated at the Ga site in GaN, and the Eu$^{2+}$/Eu$^{3+}$ ratio is dependent on the position of Fermi level and thus the growth conditions. We have discussed the tuning of the Eu valence state and concentration in terms of {\it global} and {\it local} effects caused by indirect and direct defect--defect interactions through co-doping and defect association, respectively. Based on a detailed analysis of the defects' local lattice environment and electronic behavior, the unassociated Eu$_{\rm Ga}$ is identified as an optically active center for sharp red Eu$^{3+}$ intra-$f$ luminescence. Eu-related defect complexes such as Eu$_{\rm Ga}$-O$_{\rm N}$, Eu$_{\rm Ga}$-Si$_{\rm Ga}$, Eu$_{\rm Ga}$-H$_i$, Eu$_{\rm Ga}$-Mg$_{\rm Ga}$, and Eu$_{\rm Ga}$-O$_{\rm N}$-Mg$_{\rm Ga}$ are expected to be more efficient for non-resonant excitation of Eu$^{3+}$. whereas Eu$_{\rm Ga}$-$V_{\rm Ga}$ is unlikely to be efficient. Eu$_{\rm Ga}$-$V_{\rm N}$ is likely to have a limited role in intra-$f$ high-energy luminescence. Eu-related defects can also act as carrier traps for defect-to-band transitions that emit visible light. The unassociated Eu$_{\rm Ga}$, for example, can be a source of the broad blue emission observed in n-type, Eu$^{2+}$-containing GaN.   

\begin{acknowledgments}

This work made use of resources in the Center for Computationally Assisted Science and Technology (CCAST) at North Dakota State University. 

\end{acknowledgments}

\appendix

\section{Effects of the semicore Ga $3d$ electrons}

\begin{figure}[h]
\centering
\includegraphics[width=0.99\linewidth]{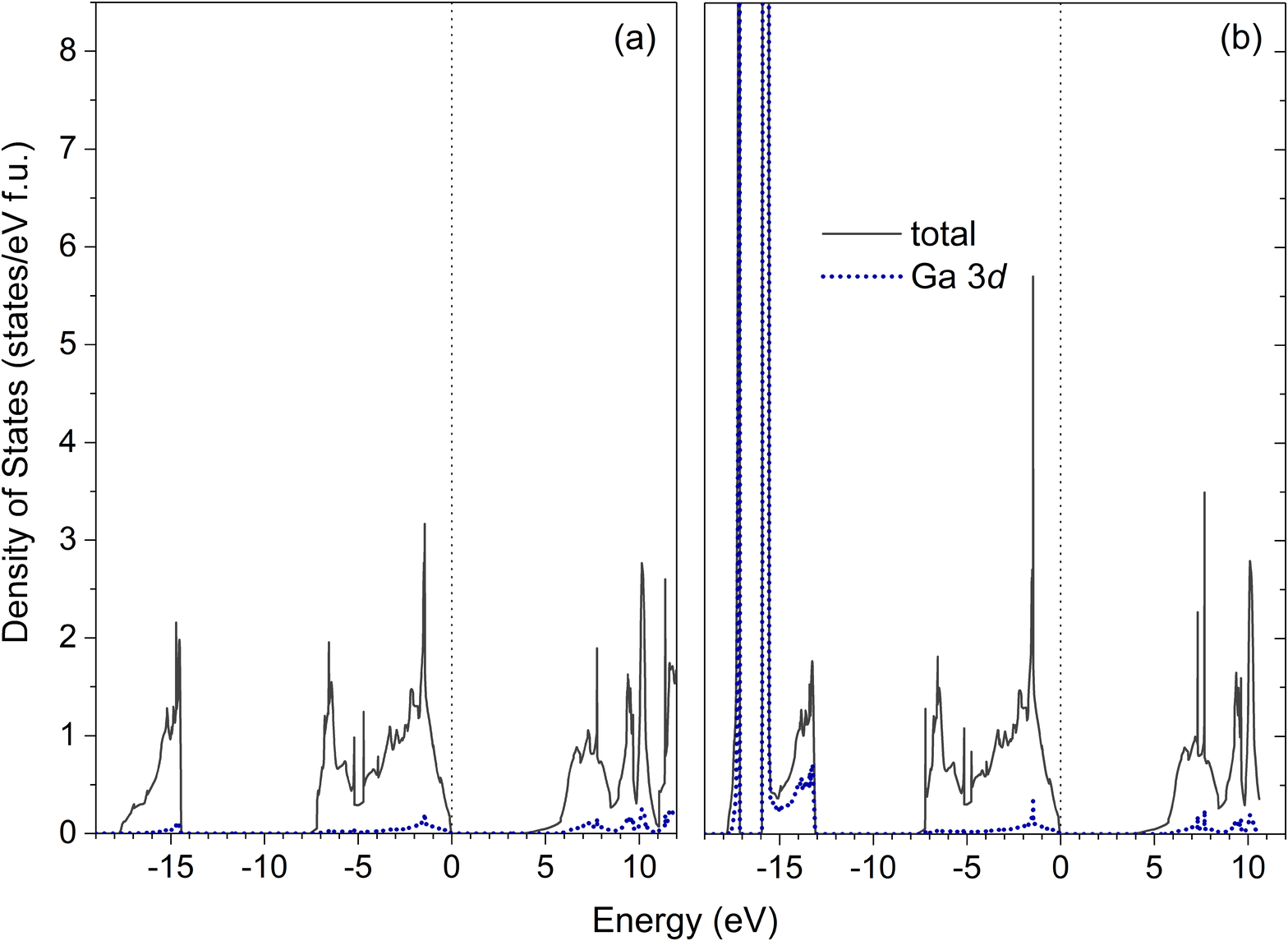}
\caption{Total and projected densities of states of bulk GaN obtained in calculations where the Ga $3d$ electrons are (a) treated as core states or (b) explicitly included in the valence. The zero of energy is set to the highest occupied state.}
\label{fig;dos;undoped}
\end{figure}

Figure \ref{fig;dos;undoped} shows the total and Ga $3d$-projected electronic densities of states (DOS) of bulk GaN in calculations using the Heyd-Scuseria-Ernzerhof (HSE) functional \cite{heyd:8207}, in which the semicore Ga $3d$ electrons are either (a) treated as core states or (b) explicitly included in the valence. The Ga $3d$ states are found to be deep in the valence band ($>$15 eV below the VBM). The inclusion of the Ga $3d$ electrons as valence electrons slightly changes the lattice parameters and band gap value. To match the experimental band gap, we set the Hartree-Fock mixing parameter ($\alpha$) to 0.310 in the Ga $3d$-as-core calculation and 0.295 in the Ga $3d$-as-valence calculation. The former gives $a=3.1964$ {\AA} and $c=5.1883$ {\AA}; the latter gives $a=3.1819$ {\AA} and $c=5.1693$ {\AA}; here, the atomic structure of GaN in the two cases is fully optimized within the HSE functional with the respective $\alpha$ values.

Our tests for Eu$_{\rm Ga}$ show that its $\epsilon(0/-)$ level is 0.55 eV below the CBM in the Ga $3d$-as-valence HSE calculations, which is 0.11 eV greater than that obtained in the Ga $3d$-as-core calculations. The difference is within the typical error bar in our calculations (expected to be about 0.1 eV for singly charged defects). The effects of the semicore electrons thus do not affect the physics of what we are presenting and our conclusions. Note that the inclusion of the Ga $3d$ electrons in the valence (which results in a 125\% increase in the total number of valence electrons in the supercell) is computationally much more demanding, making it prohibitive to carry out calculations for a large number of defect configurations.              

\section{Electronic structure of Eu- {\it versus} Er-doped GaN}

\begin{figure}[h]
\centering
\includegraphics[width=0.99\linewidth]{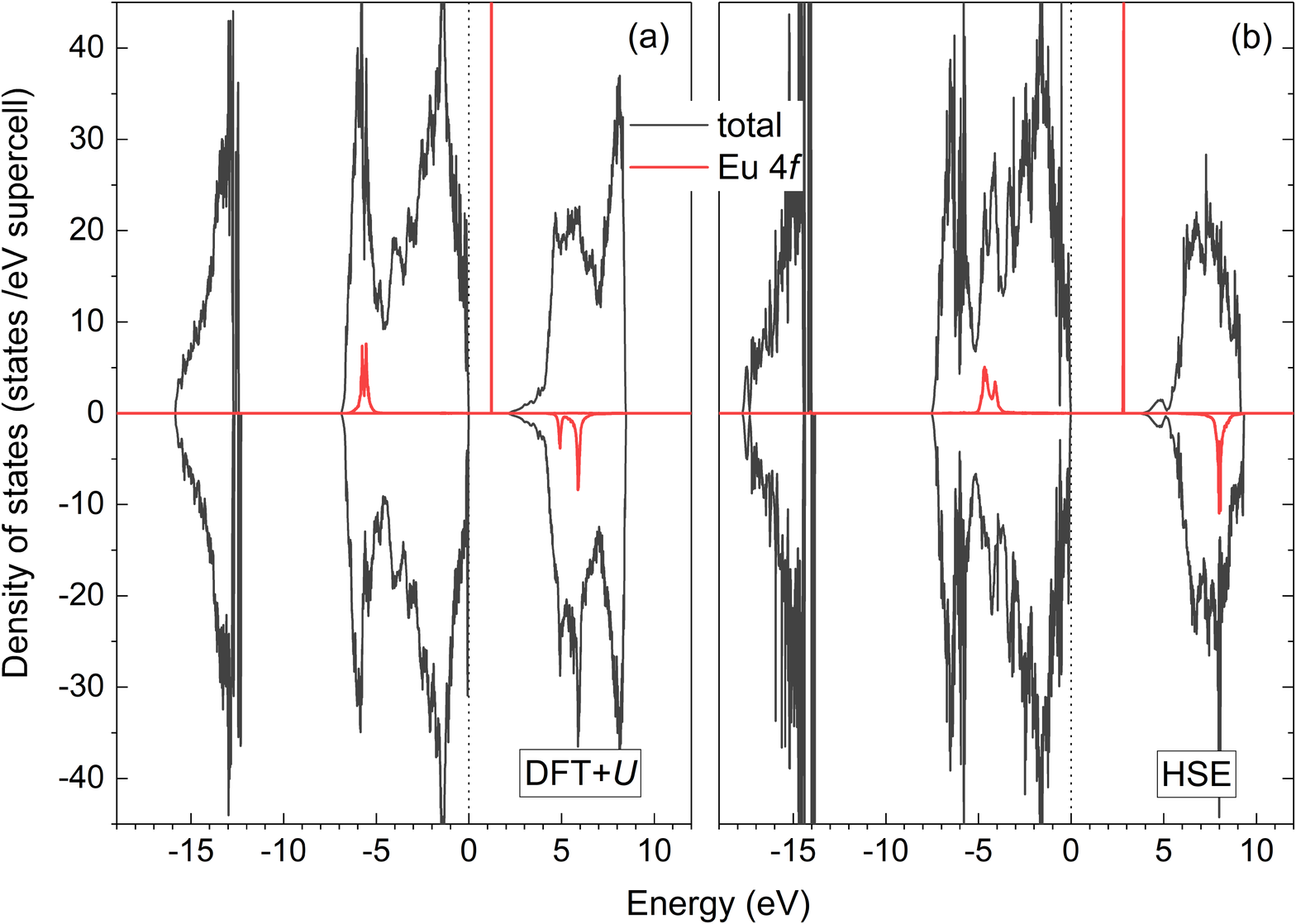}
\caption{Total and projected densities of states of Eu-doped GaN obtained in calculations using (a) the DFT$+$$U$ method ($U^{\rm eff} = 7$ eV for Eu $4f$ electrons) or (b) the HSE functional. The zero of energy is set to the highest occupied state.}
\label{fig;dos;eudoped}
\end{figure}

Figure \ref{fig;dos;eudoped} show the total and Eu $4f$-projected electronic densities of states of Eu-doped GaN obtained in DFT$+$$U$ \cite{Dudarev1998} and HSE \cite{heyd:8207}. In both methods, the DFT part is based on the Perdew-Burke-Ernzerhof (PBE) parametrization \cite{GGA} of the generalized gradient approximation (GGA). In these calculations, one Ga atom in the 96-atom GaN supercell is substituted with europium (Eu); i.e., the chemical composition is EuGa$_{47}$N$_{48}$ (i.e., the Eu concentration $\sim$2\%), which corresponds to the Eu$_{\rm Ga}^0$ defect configuration (i.e., Eu$^{3+}$ at the Ga$^{3+}$ lattice site) discussed in the main text. The ground state of Eu$^{3+}$ is $4f^{6}$. In the calculated DOS, we find that Eu has 6 spin-up occupied $4f$ states in the valence band, 1 spin-up unoccupied $4f$ state in the host band gap, and 7 spin-down unoccupied $4f$ states in the conduction band; see Fig.~\ref{fig;dos;eudoped}. The presence of the Eu $4f$ state in the band gap is key to the stability of Eu$^{2+}$ in Eu-doped GaN. An electron when added to the system will occupy the lowest unoccupied state, which is, in this case, the unoccupied in-gap Eu $4f$ state. Upon capturing the electron, Eu$^{3+}$ ($4f^{6}$) becomes Eu$^{2+}$ ($4f^{7}$); or, in our defect notation, Eu$_{\rm Ga}^0$ becomes Eu$_{\rm Ga}^-$. The main differences between the DFT$+$$U$ and HSE results are (i) in the calculated value of the host band gap and thus the position of the in-gap Eu $4f$ state with respect to the band edges, and (ii) the position of the Eu $4f$ levels in the valence and conduction bands.

\begin{figure}[h]
\centering
\includegraphics[width=0.99\linewidth]{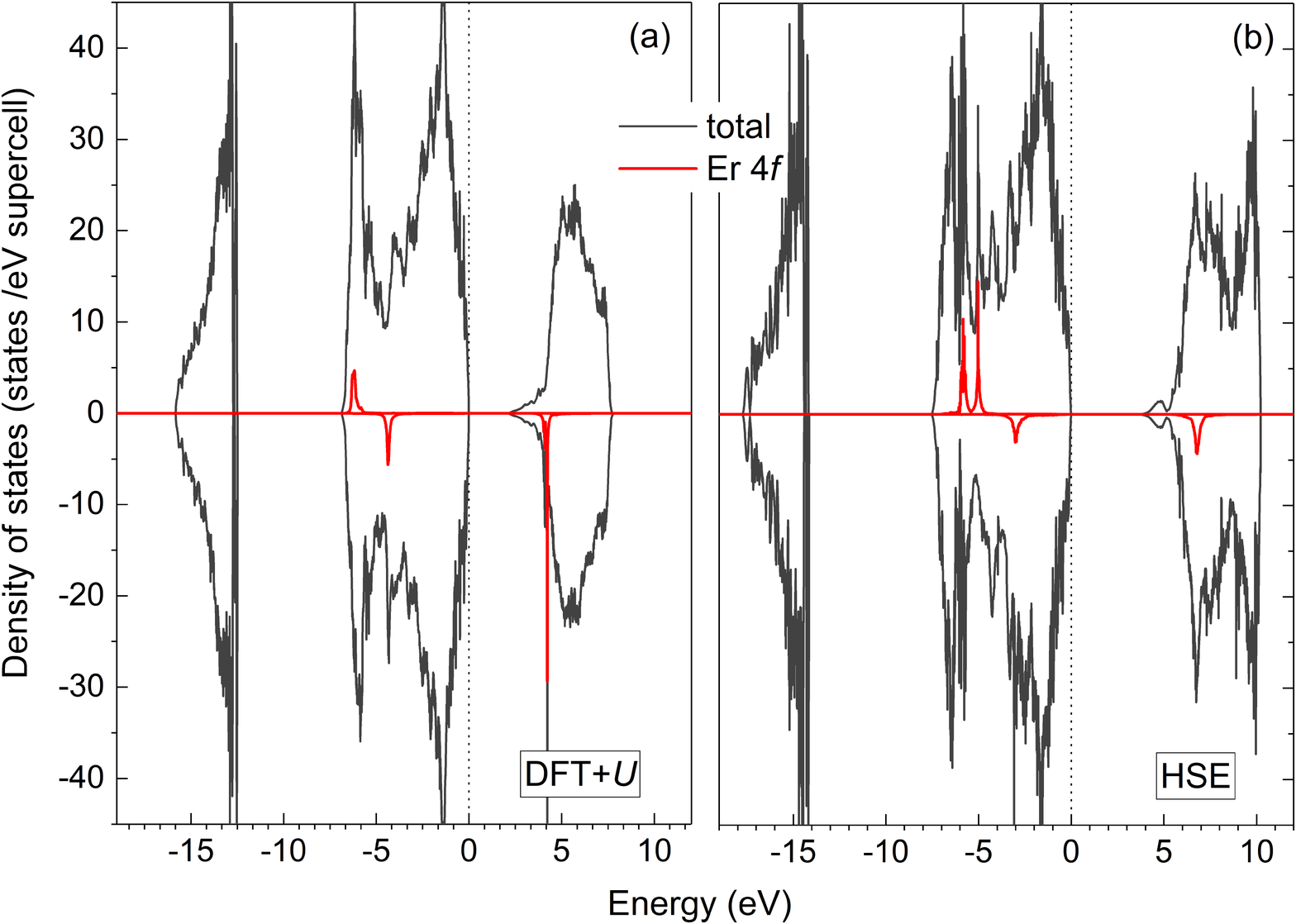}
\caption{Total and projected DOS of Er-doped GaN obtained in calculations using (a) the DFT$+$$U$ method ($U^{\rm eff} = U -J = 8$ eV for Er $4f$ electrons) or (b) the HSE functional. The zero of energy is set to the highest occupied state.}
\label{fig;dos;erdoped}
\end{figure}

For comparison, we also perform similar calculations for erbium (Er) doped GaN. Figure \ref{fig;dos;erdoped} show the total and Er $4f$-projected electronic densities of states of Er-doped GaN obtained in DFT$+$$U$ and HSE calculations. In these calculations, the chemical composition is ErGa$_{47}$N$_{48}$, which corresponds to the Er$_{\rm Ga}^0$ defect configuration (i.e., Er$^{3+}$ at the Ga$^{3+}$ site) reported in Ref.~\cite{Hoang2015RRL}. The ground state of Er$^{3+}$ is $4f^{11}$. In the calculated DOS, Er has 7 spin-up and 4 spin-down occupied $4f$ states in the valence band and 3 spin-down unoccupied $4f$ states in the conduction band. There are no Er $4f$ states in the band gap or at the band edges of the GaN host. As a result, Er is stable only as Er$^{3+}$ and the unassociated Er$_{\rm Ga}$ is electrically inert as reported in our previous work \cite{Hoang2015RRL}. 

\section{Charge density associated with Eu$_{\rm Ga}^-$}

\begin{figure}[h]
\centering
\includegraphics[width=0.55\linewidth]{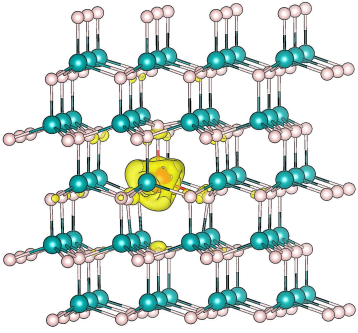}
\caption{Structure and charge density associated with Eu$_{\rm Ga}^-$ in GaN. The isovalue for the charge-density isosurface (yellow) is set to 0.02 e/{\AA}$^3$. The large (red) sphere is Eu, medium (blue) spheres are Ga, and small (red) spheres are N.}
\label{fig;chgdiff}
\end{figure}

Figure \ref{fig;chgdiff} shows the charge density associated with the extra electron in Eu$_{\rm Ga}^-$, compared to the neutral defect Eu$_{\rm Ga}^0$ but in the Eu$_{\rm Ga}^-$ atomic configuration. The electron is highly localized at the Eu site. An examination of the orbital-projected wavefunctions show that the extra electron occupies the previously unoccupied Eu $4f$ state in the band gap (see Fig.~\ref{fig;dos;eudoped}). Thus, upon capturing the electron, the defect configuration Eu$_{\rm Ga}^0$ (i.e., Eu$^{3+}$) becomes Eu$_{\rm Ga}^0$ (Eu$^{2+}$), as discussed in the main text.

\section{Structure of (Eu$_{\rm Ga}$-$V_{\rm N}$)$^{2+}$: basal {\it versus} axial}

\begin{figure}[h]
\centering
\includegraphics[width=0.99\linewidth]{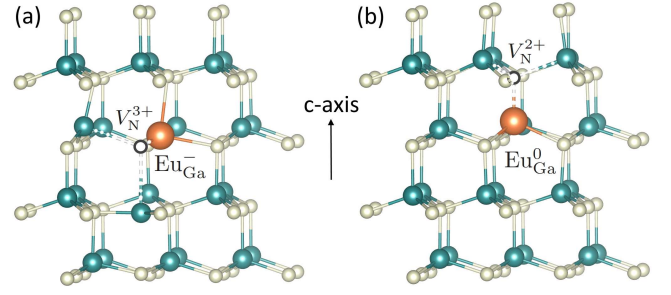}
\caption{Structure of (Eu$_{\rm Ga}$-$V_{\rm N}$)$^{2+}$: (a) basal and (b) axial geometric configurations. Large spheres are Eu, medium Ga, small N. The nitrogen vacancy is represented by a black circle.}
\label{fig;eugavn}
\end{figure}

Figure \ref{fig;eugavn} shows the structure of (Eu$_{\rm Ga}$-$V_{\rm N}$)$^{2+}$. In the basal geometric configuration, Eu is found to be stable as Eu$^{2+}$, and (Eu$_{\rm Ga}$-$V_{\rm N}$)$^{2+}$ a complex of Eu$_{\rm Ga}^-$ and $V_{\rm N}^{3+}$; the Eu$^{2+}$ is off-center by 0.58 {\AA}. This defect configuration is 83 meV lower in energy than another basal configuration where Eu is stable as Eu$^{3+}$ and (Eu$_{\rm Ga}$-$V_{\rm N}$)$^{2+}$ is a complex of Eu$_{\rm Ga}^0$ and $V_{\rm N}^{2+}$. In the axial geometric configuration, Eu is stable as Eu$^{3+}$, and (Eu$_{\rm Ga}$-$V_{\rm N}$)$^{2+}$ is a complex of Eu$_{\rm Ga}^0$ and $V_{\rm N}^{2+}$; the Eu$^{3+}$ is off-center by only 0.23 {\AA}. Here, Eu cannot be stabilized as Eu$^{2+}$ in the axial configuration. The difference between the two geometric configurations of (Eu$_{\rm Ga}$-$V_{\rm N}$)$^{2+}$ can be understood in terms of their local lattice environment. In this case, it is energetically less unfavorable for the basal configuration to accommodate the larger Eu$^{2+}$ ion by moving the ion significantly off-center. The strong Coulomb interaction between Eu$_{\rm Ga}^-$ and $V_{\rm N}^{3+}$ helps significantly lower the formation energy of that particular defect complex configuration, making it energetically more favorable than the alternative configuration involving Eu$^{3+}$. Also note that, in the unassociated $V_{\rm N}$, $V_{\rm N}^{2+}$ is energetically less stable than either $V_{\rm N}^{+}$ or $V_{\rm N}^{3+}$ in the entire range of the Fermi-level values from the VBM to the CBM \cite{Yan2012APL,Hoang2015RRL,Lyons2017NPJCM}.

\section{Comparison with previous calculations}

In response to the reviewers' request, we provide in the following a brief summary of relevant computational studies of Eu-doped GaN previously reported in the literature and a comparison with our current work.

Filhol et al.~\cite{Filhol2004} investigated rare-earth (RE) dopants in wurtzite GaN based on DFT within the spin-polarized local-density approximation (LDA). In their calculations, the $4f$ electrons of the RE (including Eu) were treated as core states; i.e., Eu, for example, was {\it a priori assumed} to be Eu$^{3+}$. Such a computational approach is clearly not suitable for the study of Eu-doped GaN since, as discussed in our work, Eu does introduce a $4f$ state in the band gap of the GaN host, and such an electronic state plays an important role in defect formation in the material. The defect energy levels for RE$_{\rm Ga}$, RE$_{\rm Ga}$-$V_{\rm N}$, and RE$_{\rm Ga}$-$V_{\rm Ga}$, and RE$_{\rm Ga}$-O$_{\rm N}$ reported by Filhol et al., obtained by examining the Kohn-Sham levels at the $\Gamma$ point, are {\it qualitatively} different from our results. For example, Eu$_{\rm Ga}$ and Eu$_{\rm Ga}$-O$_{\rm N}$ were reported to be electrically inert and have no defect energy levels in the host band gap \cite{Filhol2004}, which is in contrast to our findings. 

Svane et al.~\cite{Svane2006} studied RE impurities in GaN (and GaAs) using self-interaction corrected, spin-polarized LDA calculations. They found that Eu introduces impurity states in the host band gap; see Fig.~4 of Ref.~\cite{Svane2006}. As far as the in-gap Eu $4f$ state is concerned, that result is similar to what we observe in Fig.~\ref{fig;dos;eudoped}. The features involved the Eu $4f$ states in the valence and conduction bands are, however, significantly different from our work [even from our results obtained in DFT$+$$U$ calculations; see Fig.~\ref{fig;dos;eudoped}(a)], which can be ascribed to the different computational methods used. Most importantly, they found that the Eu$_{\rm Ga}^{0/-}$ (i.e., Eu$^{3+/2+}$) level is above the calculated CBM (i.e., Eu is thus stable only as Eu$^{3+}$ in the entire range of the Fermi-level values from the VBM to the CBM), which is {\it in contrast} to what we find and to the fact that Eu is known to be mixed valence. Note that, in their defect calculations, finite-supercell-size effects were not corrected and the atomic structures were not optimized. Also note that Svane et al.~considered zincblende GaN, although both the zincblende and wurtzite structures are expected to show very similar defect physics.  

Sanna et al.~\cite{Sanna2009} adopted the LDA$+$$U$ scheme within the density-functional-based tight-binding (DFTB) method for their study of RE impurities in wurtzite GaN. The effective $U$ value, $U^{\rm eff}$, was chosen to be 6.8 eV for the Eu $4f$ states. They found the Eu$_{\rm Ga}^{0/-}$ level to be at 1.58 eV above the VBM, much lower than our result ($E_v + 3.09$ eV). Note that the authors did not discuss the stability of Eu$^{2+}$. If we assume that their Eu$_{\rm Ga}^-$ corresponds to Eu$^{2+}$, then the result means that Eu$^{2+}$ is stable in a large range of Fermi-level values below the CBM [even larger than the Fermi-level range where Eu$^{3+}$, i.e., Eu$_{\rm Ga}^0$, is more stable]. Such a result is not consistent with what is known experimentally according to which the Eu$^{2+}$/Eu$^{3+}$ ratio is low even under n-type conditions, except when Eu-doped GaN is co-doped with both O and Si \cite{Nunokawa2020JAP}.

The Eu$_{\rm Ga}$-$V_{\rm N}$ complex were reported to have the $(+/0)$ level at 0.27 eV and the $(0/-)$ at 3.14 eV above the VBM; and Eu$_{\rm Ga}$-$V_{\rm Ga}$ to have the $(0/-)$ level at 0.5 eV, the $(-/2-)$ level at 0.7 eV, the $(2-/3-)$ level at 1.1 eV, and the $(3-/4-)$ level at 2.59 eV above VBM \cite{Sanna2009}. These results are {\it qualitatively} different from those reported in our work. For example, Sanna et al.~did not find the $(3+/2+)$ and $(2+/+)$ levels of Eu$_{\rm Ga}$-$V_{\rm N}$. These defect levels should be expected to be stable given that $V_{\rm N}$ is stable predominantly in positively charged states in GaN. Note that even for the isolated $V_{\rm N}$, the most well studied native point defect in GaN, their results (see Fig.~5 of Ref.~\cite{Sanna2009}) were also very different from those obtained in calculations using more advanced methods \cite{Yan2012APL,Hoang2015RRL,Lyons2017NPJCM}. The discrepancy can be ascribed mainly to the use of DFTB in the previous work which apparently was not able to provide an accurate description of defect physics in GaN.

Ouma et al.~\cite{Ouma2014PhysicaB} investigated the Eu$_{\rm Ga}$-$V_{\rm N}$ complex in wurtzite GaN using DFT calculations within the generalized gradient approximation (GGA). In their defect calculations, Eu was {\it assumed} to be stable as Eu$^{3+}$ and finite-size corrections were not included. Their results for the defect complex, with the presence of the $(-/2-)$ and $(0/2-)$ levels, are {\it qualitatively} different from ours, and not consistent with the fact that $V_{\rm N}$ is stable predominantly in positively charged states. There was no discussion of the stability of the Eu$^{2+}$ ion. 

Mitchell et al.~\cite{Mitchell2017PRB} reported preliminary results on the local lattice structure and charge density of different charge states of Eu$_{\rm Ga}$-$V_{\rm Ga}$, obtained from GGA$+$$U$ calculations ($U^{\rm eff} = 6.8$ eV for Eu $4f$ states). The authors {\it a priori assumed} the defect complex to have charge states from $0$ to $3-$ and Eu to remain in the trivalent state. 

There are studies of Eu-doped GaN that were based on an examination of the electronic density of states obtained in standard DFT (within LDA or GGA) and/or DFT$+$$U$ calculations. Goumri-Said et al.~\cite{GoumriSaid2008JPD} in LDA$+$$U$ calculations (with $U^{\rm eff} = 5 $ eV for the Eu $4f$ states), for example, found that there were no Eu $4f$ states in the band gap of the zincblende GaN host but $4f$ states in the valence and conduction bands and near the VBM and CBM, respectively; see Fig.~2 of Ref.~\cite{GoumriSaid2008JPD}. Su et al.~\cite{Su2018PhysicaB}, in GGA$+$$U$ ($U^{\rm eff} = 7.4$ eV) calculations for Eu$_{\rm Ga}^0$, also did not find in-gap Eu $4f$ states. Cruz et al.~\cite{Cruz2012PRB}, based on their GGA and GGA$+$$U$ ($U^{\rm eff} = 6$ eV) calculations for Eu$_{\rm Ga}^0$ in wurtzite GaN, also reported similar results. All these results are thus {\it in contrast} to our finding reported in Fig.~\ref{fig;dos;eudoped} and also discussed in the main text. Note that our results obtained in DFT$+$$U$ calculations with $U^{\rm eff} = 5$ eV or $6$ eV are very similar to that reported in Fig.~\ref{fig;dos;eudoped}(a) in which Eu introduces an in-gap $4f$ state. This may suggest that, in some of the previous DFT$+$$U$ calculations, the system was not fully converged to its ground state. The high Eu concentration of Eu (i.e., small supercell sizes) in some calculations may have also introduced strong spurious defect-defect interaction; as a result, the physics at the dilute limit could not be captured properly.    

Masago et al.~\cite{Masago2014JJAP,Masago2014APEx,Masago2014APE,Masago2017APE} reported the electronic structure and energetics of (Eu,Mg)- and (Eu,Mg,O)-doped GaN. Specifically, they investigated the following defect configurations using DFT calculations within GGA: Eu$_{\rm Ga}$-Eu$_{\rm Ga}$, Eu$_{\rm Ga}$-Mg$_{\rm Ga}$-Eu$_{\rm Ga}$, and Eu$_{\rm Ga}$-O$_{\rm N}$-Mg$_{\rm Ga}$-Eu$_{\rm Ga}$, all involving a pair of Eu$_{\rm Ga}$ defects. We do not consider these defect complexes in our current work as they are not relevant to the physics we are investigating. 

In short, given the known shortcomings of semilocal functionals such as LDA/GGA \cite{Freysoldt2014RMP,Dierolf2016Book}, care should be taken, however, when interpreting the results. Also note that, in addition to the issue with the electronic structure (including the so-called ``band-gap problem''), DFT and DFT$+$$U$ calculations based on LDA/GGA are known to provide a poor description of the structure and hence the energetics of defects, even simple native defects \cite{Lyons2015}.

%\bibliography{../../optoelectronics_refs}

%

\end{document}